\documentclass{aa}     % LaTeX A&A  Standard Fonts
\usepackage{graphics}

\def\cc{\,{\rm cm^{-3}}}
\def\cm2{\,{\rm cm^{-2}}}
\def\kms{\,{\rm {km\,s^{-1}}}}
\def\kkms{\,{\rm {K\,km s^{-1}}}}
\def\co{\,{\rm ^{12}CO}}
\def\13co{\,{\rm ^{13}CO}}
\def\h2{\,{\rm H_{2}}}
\def\Msun{\rm M_{\odot}}
\def\Lsun{\rm L_{\odot}}

\def\aua{{\rm A\&A} }
\def\auas{{\rm A\&AS} }

\def\apj{{\rm ApJ} }
\def\aj{{\rm AJ} }
\def\apjs{{\rm ApJS} }
\def\apjl{{\rm ApJL} }

\def\mnras{{\rm MNRAS} }
\def\pasj{{\rm PASJ} }

\begin{document}

\title{CI and CO in nearby galaxy centers}

\subtitle{The bright galaxies NGC~1068 (M~77), NGC~2146, NGC~3079, NGC~4826 (M~64), and NGC~7469}

\author{F.P. Israel \inst{1}
       }
 
   \offprints{F.P. Israel}
 
  \institute{Sterrewacht Leiden, Leiden University, P.O. Box 9513, 2300 RA 
             Leiden, The Netherlands
  }
 
\date{Received ????; accepted ????}
 
\abstract{} {We study the physical properties and amount of molecular
  gas in the central regions of galaxies with active nuclei.}  {Maps
  and measurements of the $J$=1--0, $J$=2--1, $J$=3--2, $J$=4--3
  $\co$, the $J$=1--0, $J$=2--1, and $J$=3--2 $\13co$ lines in the
  central arcminute squared of NGC~1068, NGC~2146, NGC~3079, NGC~4826,
  and NGC~7469, as well as 492 GHz CI maps in three of these are used
  to model the molecular gas clouds in these galaxies.}  {Bright CO
  concentrations were detected and mapped in all five objects.  In all
  cases, the observed lines could be fitted with two distinct gas
  components. The physical condition of the molecular gas is found to
  differ from galaxy to galaxy. Rather high kinetic temperatures of
  125-150 K occur in NGC~2146 and NGC~3079. Very high densities of
  $0.3-1.0\,\times\,10^{5}\,\cc$ occur in NGC~2146, NGC~3079, and
  NGC~7469. The CO to $\h2$ conversion factor $X$ is typically an
  order of magnitude less than the `standard' value in the Solar
  Neighborhood.  The molecular gas is constrained within radii
  between 0.9 and 1.5 kpc from the nuclei. Within these radii, $\h2$
  masses are typically $1.2-2.5\,\times\,10^{8}$ M$_{\odot}$.  The
  exception is the (relatively nearby) merger NGC~4826 with $R$=0.3
  kpc, and $M\,=\,3\times\,10^{7}$ M$_{\odot}$.  The $\h2$ mass is
  typically about one per cent of the dynamical mass in the same
  region.}  {} { 

\titlerunning{CI and CO in five bright spirals}
\maketitle

\section{Introduction}

Molecular gas is a major constituent of the interstellar medium in
galaxies.  This is particularly true for star-forming complexes in the
spiral arms, but strong concentrations of molecular gas are also
frequently found in the inner few kiloparsec of spiral galaxies.
These concentrations of gas play an important role in the evolution of
galaxy centers.  They provide the material for inner galaxy starbursts
and the accretion of massive black holes.  It is thus important to
determine the characteristics of this molecular hydrogen gas (density,
temperature, excitation) and especially its amount.  It cannot be
observed directly, and its properties can only be inferred from
observations of tracer elements, of which CO is one of the most
abundant and easiest observable.  However, the CO emitting gas is not
in LTE, and the most commonly observed $\co$ lines are optically
thick.  We have to observe CO in various transitions to obtain
reliable physical results and to break the temperature-density
degeneracy that plagues $\co$ intensities, also in an optically thin
isotope (e.g. also in $\13co$). The observed molecular and atomic line
intensities then provide the essential input for further modeling.

We have therefore observed a sample of nearby spiral galaxy centers in
various CO transitions and in the 492 GHz $^{3}$P$_{1}$--$^{3}$P$_{0}$
[CI] transition.  These galaxies were selected to be bright at
infrared wavelengths, and more specifically to have IRAS
flux densities f$_{12\mu m}\,\geq\,1.0$ Jy.  The results for seven
galaxies from this sample have already been published. These are
NGC~253 (Israel et al. 1995), NGC~7331 (Israel $\&$ Baas 1999),
NGC~6946, and M~83 = NGC~5236 (Israel $\&$ Baas 2001 -- Paper I), IC~342
and Maffei 2 (Israel $\&$ Baas 2003 -- Paper II), M~51 = NGC~5194
(Israel et al. 2006 -- Paper III).  In this paper, we present results
obtained for an additional five bright and well-studied galaxies.  In
Table\,\ref{galparm} we have summarized the characteristics of their
appearance.

%Figure1: NGC 1068
\begin{figure}[]
\unitlength1cm
\begin{minipage}[t]{8.9cm}
\resizebox{9.cm}{!}{\rotatebox{270}{\includegraphics*{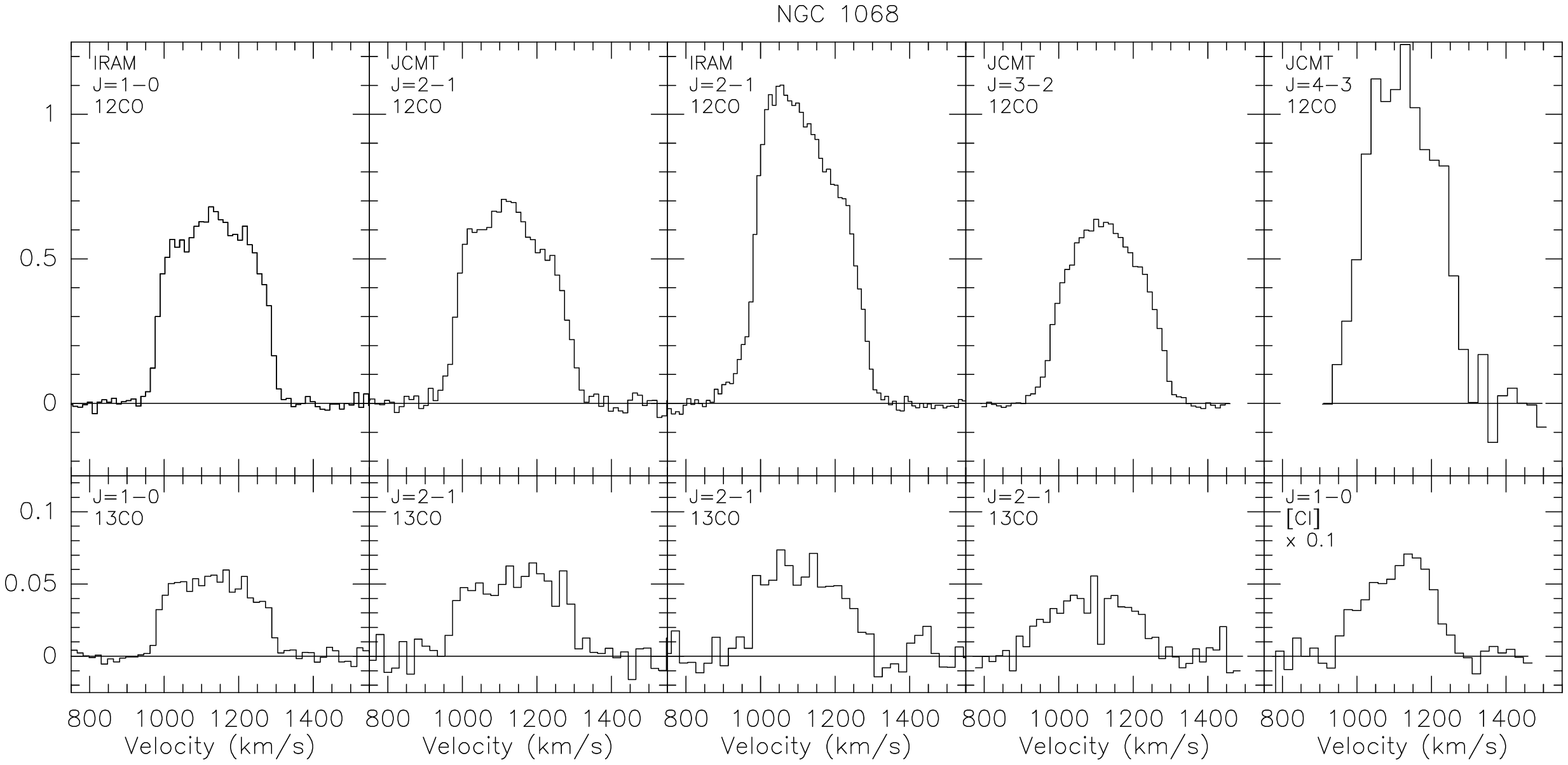}}}
\end{minipage}

\vspace{0.2cm}
\begin{minipage}[t]{8.9cm}
\resizebox{4.5cm}{!}{\rotatebox{270}{\includegraphics*{N1068_230_20Kkms.ps}}}
\resizebox{4.5cm}{!}{\rotatebox{270}{\includegraphics*{N1068_345_15Kkms.ps}}}
\end{minipage}

\vspace{0.2cm}
\begin{minipage}[t]{8.9cm}
\resizebox{9.cm}{!}{\rotatebox{270}{\includegraphics*{N1068_345pV_100mK.ps}}}
\end{minipage}
\caption[] {CO in the center of NGC~1068.  Top: Observed line
  profiles.  Horizontal scale is LSR velocity $V_{\rm LSR}$ in
  km/sec. Vertical scale is main-beam brightness temperature $T_{\rm
    mb}$ in Kelvins, using the efficiencies listed in Table~2. Note
  different scales for $^{12}$CO, $^{13}$CO and [CI] respectively.
  Center: (Left) $J$=2-1 $^{12}$CO emission integrated over the
  velocity interval 950-1350 $\kms$; contours are in steps of 20
  $\kkms$; here and in all other figures, the first contour is equal
  to the step value. (Right) $J$=3-2 $^{12}$CO emission integrated
  over the same velocity interval; contours step is 15 $\kkms$.
  Bottom: $J$=3-2 $^{12}$CO; position-velocity map in position angle
  PA = 90$^{o}$ integrated over a strip 20$''$ wide with contours in
  steps of 0.1 K; east is at top. }
\label{n1068COmaps}
\end{figure}

%Figure2: NGC2146 Maps
\begin{figure}[]
\unitlength1cm
\begin{minipage}[t]{9.cm}
\resizebox{9.cm}{!}{\rotatebox{270}{\includegraphics*{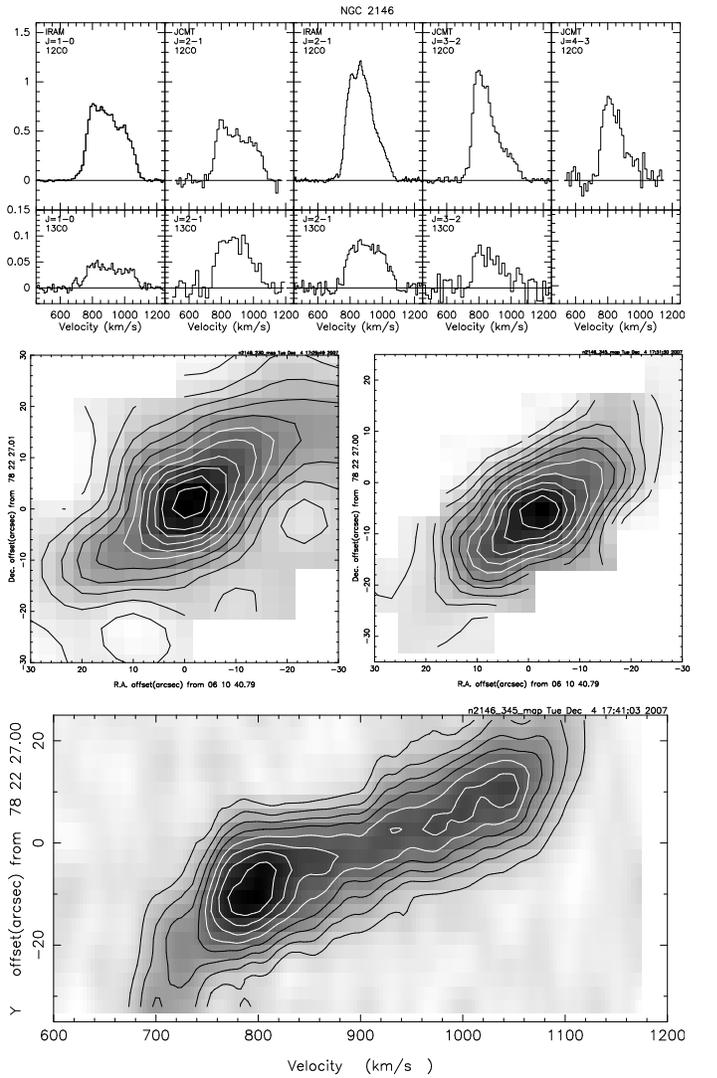}}}
\end{minipage}

\vspace{0.2cm}
\begin{minipage}[t]{9.cm}
\resizebox{4.5cm}{!}{\rotatebox{270}{\includegraphics*{N2146_230_15kkms.ps}}}
\resizebox{4.5cm}{!}{\rotatebox{270}{\includegraphics*{N2146_345_20kkms.ps}}}
\end{minipage}

\vspace{0.2cm}
\begin{minipage}[t]{9.cm}
\resizebox{9.cm}{!}{\rotatebox{270}{\includegraphics*{N2146_345pV_100mK.ps}}}
\end{minipage}
\caption[] {Center of NGC~2146, as Figure 1.  Top: Observed line
  profiles. Center: (Left) $J$=2-1 $^{12}$CO emission integrated over
  the velocity interval 600-1200 $\kms$; contours step is 15
  $\kkms$; (Right) $J$=3-2 $^{12}$CO emission integrated over the same
  velocity interval; contour step is 20 $\kkms$.  Bottom: $J$=3-2
  $^{12}$CO; position-velocity map in position angle PA = 303$^{o}$
  integrated over a strip 20$''$ wide with contours in steps of 0.1 K;
  northwest is at top. }
\label{n2146COmaps}
\end{figure}

%Figure3: NGC 3079
\begin{figure}[]
\unitlength1cm
\begin{minipage}[t]{8.9cm}
\resizebox{9.cm}{!}{\rotatebox{270}{\includegraphics*{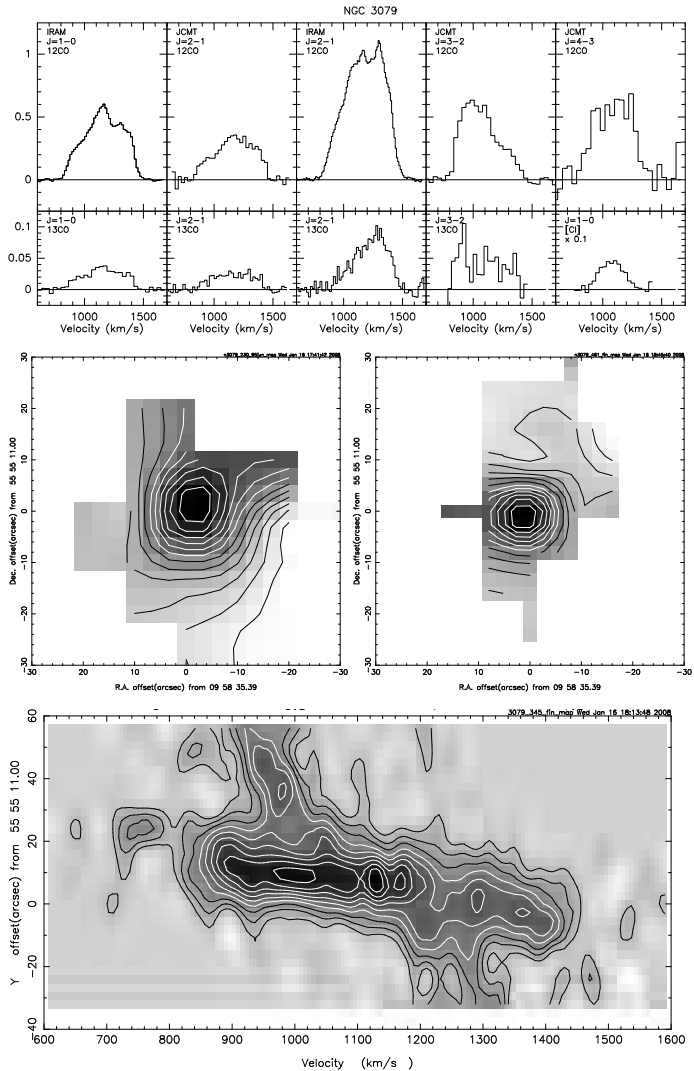}}}
\end{minipage}

\vspace{0.2cm}
\begin{minipage}[t]{8.9cm}
\resizebox{4.5cm}{!}{\rotatebox{270}{\includegraphics*{N3079_230_15kkms.ps}}}
\resizebox{4.5cm}{!}{\rotatebox{270}{\includegraphics*{N3079_461_20Kkms.ps}}}
\end{minipage}

\vspace{0.2cm}
\begin{minipage}[t]{8.9cm}
\resizebox{9.cm}{!}{\rotatebox{270}{\includegraphics*{N3079_345pV_40mK.ps}}}
\end{minipage}
\caption[] {Center o NGC~3079, as Figure 1.  Top: Observed line
  profiles. Center: (Left) $J$=2-1 $^{12}$CO emission integrated over
  the velocity interval 950-1350 $\kms$; contour step is 15 $\kkms$;
  (Right) $J$=4-3 $^{12}$CO emission integrated over the same velocity
  interval; contour step is 20 $\kkms$.  Bottom: $J$=3-2 $^{12}$CO;
  position-velocity map in position angle PA = -11$^{o}$ integrated
  over a strip 20$''$ wide with contours in steps of 0.04 K; north is
  at top. }
\label{n3079COmaps}
\end{figure}

%Figure4: NGC4826 Maps
\begin{figure}[]
\unitlength1cm
\begin{minipage}[t]{9.cm}
\resizebox{9.cm}{!}{\rotatebox{270}{\includegraphics*{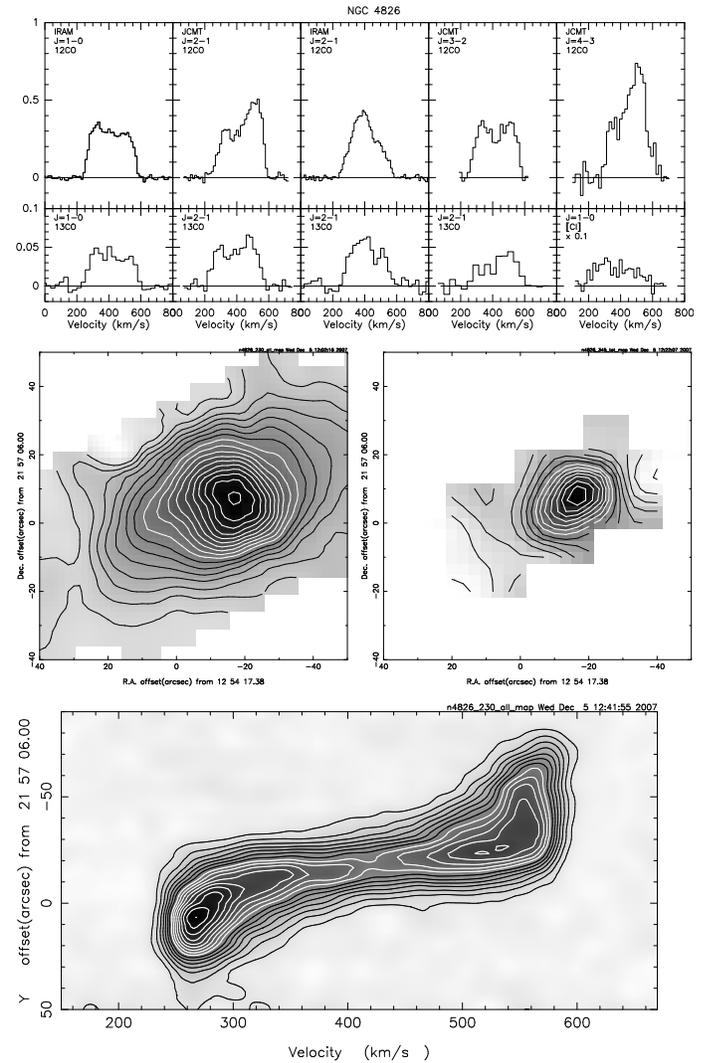}}}
\end{minipage}

\vspace{0.2cm}
\begin{minipage}[t]{9.cm}
\resizebox{4.5cm}{!}{\rotatebox{270}{\includegraphics*{N4826_230_5kkms.ps}}}
\resizebox{4.5cm}{!}{\rotatebox{270}{\includegraphics*{N4826_345_5kkms.ps}}}
\end{minipage}

\vspace{0.2cm}
\begin{minipage}[t]{9.cm}
\resizebox{9.cm}{!}{\rotatebox{270}{\includegraphics*{N4826_230pV_30mK.ps}}}
\end{minipage}
\caption[] {Center of NGC~4826, as Figure 1.  Top: Observed line
  profiles. Center: (Left) $J$=2-1 $^{12}$CO emission integrated over
  the velocity interval 150-650 $\kms$; contour step is 5 $\kkms$;
  (Right) $J$=3-2 $^{12}$CO emission integrated over the same velocity
  interval; contour step is 5 $kkms$.  Bottom: $J$=2-1 $^{12}$CO;
  Bottom: position-velocity map in position angle PA = 115${o}$
  integrated over a strip 20$''$ wide with contours in steps of 0.03
  K; northwest is at top. }
\label{n4826COmaps}
\end{figure}

%Figure5: NGC 7469
\begin{figure}[]
\unitlength1cm
\begin{minipage}[t]{8.9cm}
\resizebox{9.cm}{!}{\rotatebox{270}{\includegraphics*{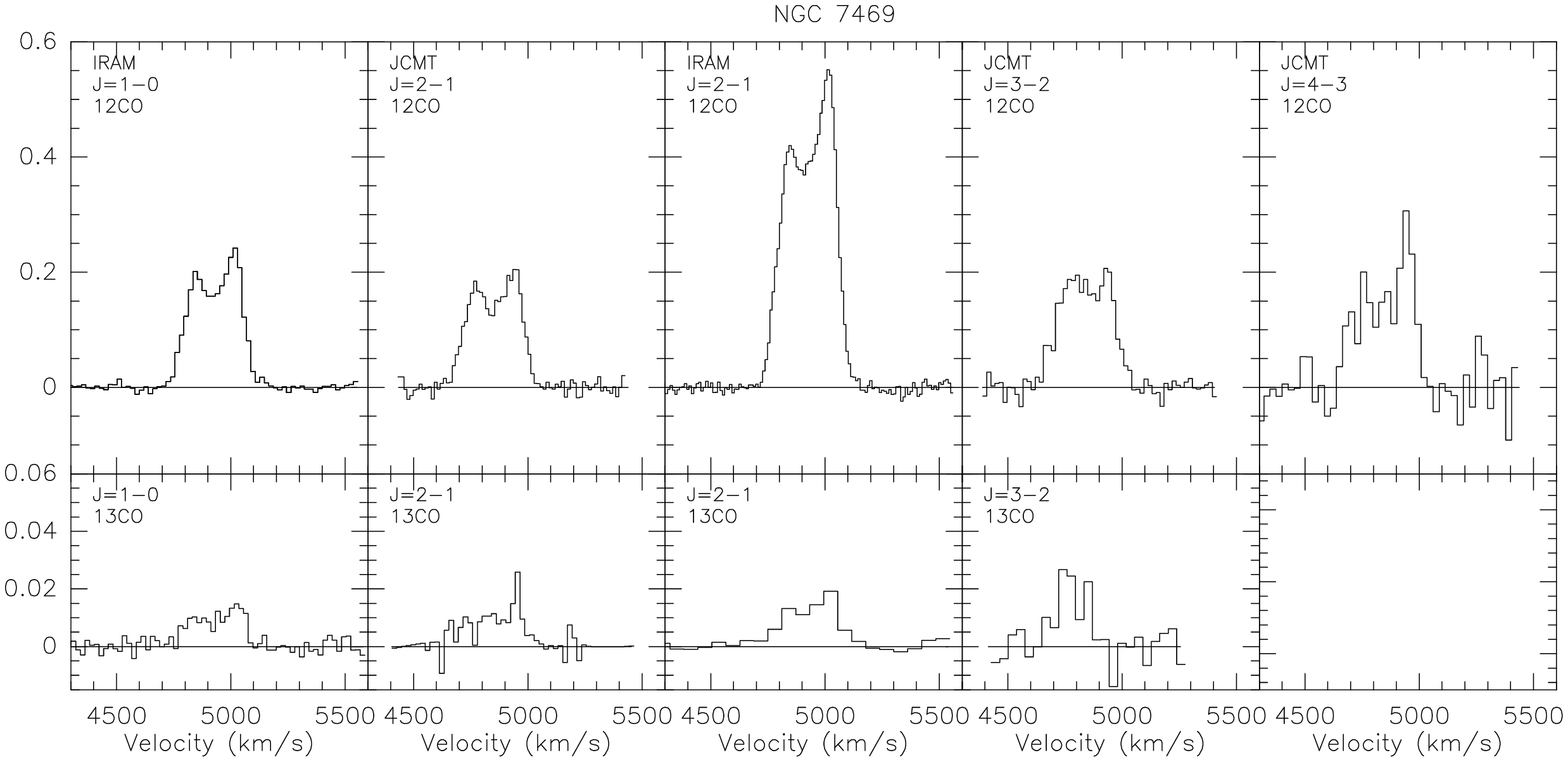}}}
\end{minipage}

\vspace{0.2cm}
\begin{minipage}[t]{8.9cm}
\resizebox{4.5cm}{!}{\rotatebox{270}{\includegraphics*{N7469_230_5Kkms.ps}}}
\resizebox{4.5cm}{!}{\rotatebox{270}{\includegraphics*{N7469_345_7.5Kkms.ps}}}
\end{minipage}

\vspace{0.2cm}
\begin{minipage}[t]{8.9cm}
\resizebox{9.cm}{!}{\rotatebox{270}{\includegraphics*{N7469_345pV_25mK.ps}}}
\end{minipage}
\caption[] {Center of NGC~7469, as Figure 1.  Top: Observed line
  profiles. Center: (Left) $J$=2-1 $^{12}$CO emission integrated over
  the velocity interval 4650-5050 $\kms$; contour step is 5
  $\kkms$; (Right) $J$=3-2 $^{12}$CO emission integrated over the same
  velocity interval; contour step is 7.5 $\kkms$.  Bottom:
  $J$=3-2 $^{12}$CO; position-velocity map in position angle PA =
  128$^{o}$ integrated over a strip 20$''$ wide with contours in steps
  of 0.025 K; east is at top. }
\label{n7469COmaps}
\end{figure}

\begin{table*}
\caption[]{Galaxy parameters}
\begin{center}
\begin{tabular}{llllll}
\hline
\noalign{\smallskip}
                   & NGC~1068		     & NGC~2146                & NGC~3079               & NGC~4826  & NGC~7469 \\
\noalign{\smallskip}
\hline
\noalign{\smallskip}
Type$^{a}$     	   & (R)SA(rs)b; Sy1	     & SB(s)abp; HII        & SB(s)c; Lin; Sy2     & (R)SA(rs)ab; HII/Lin& (R)SAB(rs)a; Sy1.2\\
R.A. (B1950)$^{a}$ &02$^{h}$40$^{m}$07.1$^{s}$&06$^{h}$10$^{m}$40.2$^{s}$& 09$^{h}$58$^{m}$35$^{s}$&12$^{h}$54$^{m}$16.3$^{s}$&23$^{h}$00$^{m}$44.4$^{s}$\\
Decl.(B1950)$^{a}$ &-00$^{\circ}$13$'$32$''$  &+78${^\circ}$22$'$29$''$ &+55$^{\circ}$55$'$16$''$&+21${^\circ}$57$'$10.4$''$&+08${^\circ}$36$'$16$''$\\
R.A. (J2000)$^{a}$ &02$^{h}$42$^{m}$40.7$^{s}$&06$^{h}$18$^{m}$37.7$^{s}$& 10$^{h}$01$^{m}$58$^{s}$& 12$^{h}$56$^{m}$43.7$^{s}$& 23$^{h}$03$^{m}$15.6$^{s}$   \\
Decl.(J2000)$^{a}$ &-00$^{\circ}$00$'$48$''$  &+78$^{\circ}$21$'$25$''$  &+55$^{\circ}$40$'$47$''$&+21${^\circ}$40$'$58$''$&+08$^{\circ}$52$'$26$''$\\
$V_{\rm LSR}^{b}$       & +1023 $\kms$    & +879 $\kms$              & +1116 $\kms$          & +408 $\kms$& +4889 $\kms$\\
Inclination $i^{b}$    & 37$^{\circ}$     & 51$^{\circ}$              & 74$^{\circ}$          & 60$^{\circ}$ & 46$^{\circ}$ \\
Position angle $P^{b}$ & 82$^{\circ}$     & 128$^{\circ}$             & 165$^{\circ}$         & 112$^{\circ}$& 128$^{\circ}$\\
Distance $D^{c}$       & 16.5 Mpc 	 & 17.9 Mpc                 & 19.4 Mpc              & 4.1 Mpc     & 29.5 Mpc\\
Scale                  & 80 pc/$''$ 	 & 87 pc/$''$               & 94 pc/$''$            & 20 pc/$''$  & 143 pc/$''$\\
\noalign{\smallskip}
\hline
\end{tabular}
\end{center}
Notes to Table 1:
$^{a}$ NED 
$^{b}$ Kaneko et al. (1989); Tarchi et al. (2004); Young et al. (1988b); Irwin \& Seaquist (1991); 
Garc\'ia-Burillo et al. (2003); Beswick, Pedlar \& McDonald (2002); Meixner et al. (1990);
Davies, Tacconi \& Genzel  (2004).
$^{c}$ see Moustakas et al. (2006)
\label{galparm}
\end{table*}

%Table 2 Log of Observations
\begin{table*}
\caption[]{$\co$ observations log}
\begin{center}
\begin{tabular}{lccccccccc}
\hline
\noalign{\smallskip}
Galaxy  & Date    	& $T_{\rm sys}$ & Beam 	& $\eta _{\rm mb}$ & t(int) & \multicolumn{4}{c}{Map Parameters} \\
        &                &              & Size   &                 &        & No.  & Size       & Spacing & P.A. \\
        &  	        & (K)	    &($\arcsec$)& 	       	  & (sec)  &pnts  &($\arcsec$)&($\arcsec$)& ($^{\circ}$)\\
\noalign{\smallskip}
\hline 
\noalign{\smallskip}
\multicolumn{6}{l}{$\co J$=1-0 (115 GHz)}\\
NGC~1068& 07Jan & 280  	      & 22   & 0.74 &  750 &  1 & --           & -- & -- \\
NGC~2146& 05Feb & 315  	      &\vline&\vline& 1200 &  1 & --           & -- & -- \\
NGC~3079& 05Oct & 264  	      &\vline&\vline& 1680 &  1 & --           & -- & -- \\
NGC~4826& 06Jul & 222  	      &\vline&\vline&  840 &  1 & --           & -- & -- \\
NGC~7469& 05Jan & 157  	      &\vline&\vline&  960 &  1 & --           & -- & -- \\
\noalign{\smallskip}
\hline 
\noalign{\smallskip}
\multicolumn{6}{l}{$\co J$=2-1 (230 GHz)}\\
NGC~1068& 07Jan & 474         & 12 & 0.53 &  750 &  1 & --           & -- & -- \\
        & 94Jan & 350         & 21 & 0.69 &  120 & 71 & 80$\times$90 &  7 & 11\\
        & 96Jan & 850         & 21 & 0.69 &  120 & 54 & 50$\times$90 & 10 & 46\\
NGC~2146& 05Feb & 290  	      & 12 & 0.53 &  480 &  1 & --           & -- & -- \\
        & 91Apr/93Apr&1420/372& 21 & 0.69&320/300& 39 & 50$\times$80 & 10 &123\\
NGC~3079& 05Oct & 352  	      & 12 & 0.53 & 1200 &  1 & --           & -- & -- \\
        & 95Jun & 585         & 21 & 0.69 &  240 & 14 & 30$\times$30 & 10 &165\\
NGC~4826& 06Jul & 331  	      & 12 & 0.53 &  780 &  1 & --           & -- & -- \\
        & 93May/94Jan&570/468 & 21 & 0.69&200/300&136 &100$\times$80 &  7 &110\\
NGC~7469& 05Jan & 224  	      & 12 & 0.53 &  960 &  1 & --           & -- & -- \\
        & 91Sep/00Oct&194/308 & 21 & 0.69&400/900&  9 & 30$\times$30 & 10 & 35\\
\noalign{\smallskip}
\hline 
\noalign{\smallskip}
\multicolumn{6}{l}{$\co J$=3-2 (345 GHz)}\\
NGC~1068& 94jan & 1180        & 14   & 0.58 &  600 & 71 & 70$\times$80 &  7 & 45\\
        & 96Jan & 1980        &\vline& 0.58 &  120 & 63 & 30$\times$63 &  6 & 45\\
NGC~2146& 96Jan & 1139        &\vline& 0.59 &  180 & 28 & 24$\times$72 &  8 &123\\
NGC~3079& 93Dec & 1293	      &\vline& 0.56 &  400 & 12 & 14$\times$96 &  8 &165\\
NGC~4826&93May/94Jan&517/939  &\vline& 0.53&200/480& 20 & 60$\times$30 & 10 &114\\
NGC~7469&96Jul/01Jun&1358/715 &\vline& 0.61&480/600& 32 & 35$\times$28 &  7 &  0\\
\noalign{\smallskip}
\hline 
\noalign{\smallskip}
\multicolumn{6}{l}{$\co J$=4-3 (461 GHz)}\\
NGC~1068& 96Jul & 3365        & 11   & 0.50 &  600 & 23 & 24$\times$24 &  6 & 45\\
NGC~2146& 01Nov & 4743        &\vline& 0.50 & 1440 &  4 & 18$\times$18 &  6 &123\\
NGC~3079&93Dec/94Mar&3360/5510&\vline& 0.50 &  600 & 13 & 24$\times$56 &  8 &165\\
NGC~4826&93Dec/94Mar&2045/4630&\vline& 0.50&320/600& 16 & 30$\times$60 & 10 &114\\
NGC~7469& 99Jul & 2008        &\vline& 0.52 & 1200 &  9 & 18$\times$18 &  6 &  0\\
\noalign{\smallskip}
\hline
\end{tabular}
\end{center}
\label{gal12colog}
\end{table*}

%Table 3 Log of 13CO Observations
\begin{table}[t]
\caption[]{$\13co$ and [CI] observations log}
\begin{center}
\begin{tabular}{lccccc}
\hline
\noalign{\smallskip}
Galaxy  & Date    	& $T_{\rm sys}$ & Beam 	& $\eta _{\rm mb}$ & t(int) \\
        &               &              & Size   &                 &        \\
        &  	        & (K)	    &($\arcsec$)& 	       	  & (sec)  \\
\noalign{\smallskip}
\hline 
\noalign{\smallskip}
\multicolumn{6}{l}{$\13co\,J$=1-0 (110 GHz)}\\
NGC~1068& 07Jan &  160 & 23   & 0.75 & 1325 \\
NGC~2146& 05Feb &  153 &\vline&\vline& 1200 \\
NGC~3079& 05Oct &  149 &\vline&\vline& 1680 \\
NGC~4826& 06Jul &  160 &\vline&\vline& 1620 \\
NGC~7469& 05Jan &  121 &\vline&\vline& 5760 \\
\multicolumn{6}{l}{$\13co\,J$=2-1 (220 GHz)}\\
NGC~1068& 96Jan &  456 & 22 & 0.69 & 3600 \\
        & 07Jan &  212 & 13 & 0.55 & 1320 \\
NGC~2146& 91Sep & 1304 & 22 & 0.69 & 3560 \\
        & 05Feb &  374 & 13 & 0.55 &  960 \\
NGC~3079& 94Mar &  370 & 22 & 0.69 & 5220 \\
        & 05Oct &  374 & 13 & 0.55 & 1200 \\
NGC~4826& 93Dec &  534 & 22 & 0.69 & 4000 \\
        & 06Jul &  203 & 13 & 0.55 &  780 \\
NGC~7469& 95Jun &  302 & 22 & 0.69 & 4200 \\
        & 05Jan &  203 & 13 & 0.55 & 1920 \\
\multicolumn{6}{l}{$\13co\,J$=3-2 (330 GHz)}\\
NGC~1068& 97Nov &  427 & 14   & 0.59 & 2400 \\
NGC~2146& 01Jan &  914 &\vline& 0.59 & 3600 \\
NGC~3079& 95Apr & 2637 &\vline& 0.58 & 1860 \\
NGC~4826& 94Apr & 1422 &\vline& 0.53 & 6360 \\
NGC~7469& 93Apr &  882 &\vline& 0.55 & 6000 \\
\noalign{\smallskip}
\hline
\noalign{\smallskip}
\multicolumn{6}{l}{[CI] $^{3}P_{1}-^{3}P_{0}$ (492 GHz)} \\
NGC~1068$^{a}$& 96Jul & 3048 & 11   & 0.50 &  800 \\
NGC~3079$^{b}$& 94Mar & 6239 &\vline& 0.50 & 1200 \\
NGC~4826$^{c}$& 94Dec & 2904 &\vline& 0.51 & 2400 \\
              & 97Mar & 3042 &\vline& 0.51 & 1200 \\
\noalign{\smallskip}
\hline
\end{tabular}
\end{center}
\label{gal13colog}
Notes:  $^{a}$ Map of 27 points covering  $24''\times48''$ with spacing $8''$ in PA = 70$^{\circ}$.
$^{b}$ Strip map in declination, 7 points covering $8''\times56''$ with spacing $8''$ in PA = 165$^{\circ}$.
$^{c}$ Small 5-point map covering $18''\times18''$ with spacing $6''$ in PA = 114$^{\circ}$.
\end{table}

At least three of the galaxies discussed in this paper have an active
(Seyfert) nucleus (NGC~1068, NGC~3079, and NGC~7469), and probably
also NGC~4826.  Two of the galaxies (NGC~2146 and NGC~4826) show
evidence of a merger event and all -- including most of the galaxies
published earlier (IC~342, Maffei 2, M~83, and NGC~6946) -- show signs
of enhanced star formation activity in the recent past or present.
NGC~4826 is as nearby as IC~342, Maffei 2, M~83, and NGC~6946)
affording similar linear resolutions of 200--300 pc.  The other
galaxies are more distant so that our angular resolution of
$10''-14''$ corresponds to a linear resolution no better than 800-2000
pc.  Such low linear resolutions do not reveal much structural detail
in the central CO distributions; millimeter array observations are
required for this.  However, the multi-transition observations of
$\co$ and $\13co$ presented here allow us to characterize the overall
physical condition of the central molecular gas in ways that arer not
possible otherwise.

\section{Sample galaxies}

\subsection{NGC~1068 = Arp~37 = M~77}

This is a well-known Seyfert galaxy, relatively luminous at radio
wavelengths, with overall and core 1.5 GHz flux densities of 3.8 Jy
and 1.1 Jy, respectively (Condon et al. 1990).  It has been mapped in
HI by Brinks et al. (1997, VLA, $8''$), in the $J$=1-0 $\co$
transition by Scoville et al. (1983; FCRAO, $45''$), Young et
al. (1995; FCRAO, $45''$), Kaneko et al. (1989; Nobeyama, $17''$),
Helfer et al. (2003, BIMA, $9''\times6''$), in the $J$=1-0 and $J$=2-1
$\co$ transitions by Planesas et al. (1989; IRAM, $22'',13''$) as well
as in the $J$=2-1 and $J$=3-2 transitions of both $\co$ and $\13co$ by
Papadopoulos $\&$ Seaquist (1999; JCMT, $22'',14''$). All maps covered
more or less the same area that we have mapped, and show only limited
detail.  The presence of significant structure at angular sizes
substantially smaller than these beam sizes was first suggested by
Scoville et al (1983), using a daring deconvolution scheme.  This was
later confirmed by the $J$=1-0 $\co$ millimeter array maps made at
much higher angular resolutions by Planesas et al. (1991, OVRO,
$\approx3''$), Kaneko et al. (1992, NMA, $\approx5''$) and in
particular the $J$=1-0 $\co$ and $\13co$ maps by Helfer $\&$ Blitz
(1995, BIMA, $\approx4''$) and the $J$=1-0 and $J$=2-1 $\co$ maps by
Schinnerer et al. (2000, IRAM, $0.7''$).  These show a compact (size
about $5''$ or 400 pc) circumnuclear source of molecular line
emission, and bright emission extending to about $15''$ (1200 pc) from
the nucleus.  In lower-resolution maps this emission fortuitously
resembles a ring, but higher resolutions clearly reveal CO-bright
spiral arms ({\it cf} Helfer $\&$ Blitz 1995).  The arms dominate the $\co$
emission, but only in the lower $J$ transitions.  The interferometer
map by Helfer $\&$ Blitz (1995) shows strong $J$=1-0 $\13co$ emission
from the arms, but nothing from the nucleus.  The compact nucleus
cannot be unambiguously identified in the $J$=3-2 $\13co$ map measured
by Papadopoulos $\&$ Seaquist (1999) either.  The nucleus is also
dominant in the light of other molecules ({\it cf} Jackson et al. 1993,
Tacconi et al. 1994; Helfer $\&$ Blitz 1995; Usero et al. 2004) that
trace high densities or high excitation.

\subsection{NGC~2146}

NGC~2146 has an optically disturbed appearance, characterized by a
high-surface brightness core crossed by a prominent dust lane, and a
number of weaker extraplanar arms and tidal features (see Sandage $\&$
Bedke, 1994).  This bright core measures several kiloparsecs in
diameter, and is a strong source of radio emission (1.4 GHz flux
density $\approx$1 Jy , see de Bruyn 1977; Condon et al. 1996; Braun
et al. 2007) against which HI is seen in absorption (Taramopoulos et
al. 2001, Tarchi et al. 2004).  Taramopoulos et al. (2001) have
presented a VLA HI map including short spacings that shows a neutral
gas filament extending from the main body of galactic HI to the south
over half a degree (160 kpc), as well as a large amount of gas
extending out of the galaxy to the north.  The HI mass of the central
region (inner few arcmin) is about $1.6\times10^{9}\,\Msun$.
Taramopoulos et al. (2001) conclude that NGC~2146 suffered a merger or
close encounter in which the intruder was shredded, about 800 million
years ago.  The bright core is the site of a powerful starburst.
Radio observations reveal numerous supernova remnants, HII regions and
water masers throughout (Kronberg $\&$ Biermann, 1981; Zhao et
al. 1996; Lisenfeld et al. 1996; Tarchi et al. 2000, 2002); a
low-luminosity AGN may be present (Inui et al. 2005).  X-ray
observations also reveal a starburst-driven superwind carrying
material away from the galaxy's plane (Armus et al. 1995; Della Ceca
et al. 1999).  It has been mapped in the $J$=1-0 $\co$ line by Young
et al. (1988b, FCRAO, $45''$; OVRO, $7''$) and Jackson $\&$ Ho (1988,
Hat Creek, $6\times5''$), in the $J$=1-0 and $J$=2-1 transitions of
both $\co$ and $\13co$ by Paglione et al (2001, FCRAO, $45''$) and Xie
et al (1994, FCRAO, $23''$) respectively, and in the $J$=3-2 $\co$
line by Dumke et al. (2001, HHT, $24''$).  The highest resolution
observations have been published by Greve et al. (2006) using both the
IRAM 30 m ($J$=1-0, $J$=2-1 $\co$; $22''$ and $13''$) and PdB array
telescopes ($J$=1-0 $\co$ and $\13co$; $J$=2-1 $\co$; 4-5$''$ and
$\approx3''$).

\subsection{NGC~3079}

This is a large and bright spiral seen almost edge-on.  Its
central region is a strong source of radio continuum emission (de
Bruyn, 1977).  It has been mapped in the radio at various resolutions
- see references in Condon et al. (1990). These authors list 1.4 GHz
flux densities of 0.84 Jy and 0.55 Jy for the whole galaxy and the
central region respectively (see also Irwin $\&$ Saikia 2003). The
center contains a nuclear source (AGN surrounded by water masers,
Trotter et al. 1998, Hagiwara et al. 2004) and two wind-driven outflow
bubbles (Veilleux et al. 1994; Cecil et al. 2001); it is a significant
X-ray source (Pietsch et al. 1998). The central region also hosts a
powerful starburst. In the plane of the galaxy, the cavity surrounding
the AGN is marked by near-infrared line emission from shocked $\h2$
extending out to 300 pc (Israel et al. 1998; see also Hawarden et
al. 1994).  At the cavity wall, hot dust reaches evaporation
temperatures (Israel et al. 1998, see also Armus et al. 1994).
Molecular gas has been mapped with a single dish in the $J$=1-0 $\co$
line by Tinney et al. (1990, NRAO, $55''$) and Young et al. (1995,
FRAO, $45''$), in the $J$=1-0 and $J$=2-1 $\co$ lines by Braine et
al. (1997), in the $J$=1-0 $\co$ and $\13co$ lines by Paglione et
al. (2001, FCRAO $45''$) and with mm arrays in the $J$=1-0 $\co$
transition by Young et al. (1988a, OVRO, $8''\times6''$), Sofue $\&$
Irwin (1992, NMA, $4''$), Tacconi et al. (1996, PdB, $2''$), Sofue et
al. (2000, NMA, $1.5''$); Koda et al. (2002, NMA, $1.5''$). HI maps have
been presented by Braun et al. (2007). [CI] emission was measured by
Gerin $\&$ Phillips (2000), and Israel $\&$ Baas (2002). Absorption
has been mapped in various lines (HI: Pedlar et al. 1996; OH: Hagiwara
et al. 2004; CH$_{3}$OH: Impellizzeri et al. 2008). All data indicate
the presence of dense molecular material extending out to almost a
kiloparsec from the nucleus.

\subsection{NGC~4826 = M~64}

This is a small and nearby early-type (Sab) galaxy. It is relatively
isolated although it seems to be part of the loose CVn I group.  Its
remarkable appearance, with a very prominent and wide dust lane
covering the northern edge of its bulge, has caused it to become known
as the `evil-eye' galaxy.  The discovery (Braun et al. 1992) that its
inner parts rotate counter to its outer parts, i.e. that its rotation
reverses sign at about $30''$ (600 pc) from the nucleus, has led to
considerable interest in this galaxy since.  This particular
kinematical signature, subsequently studied by Rubin (1994), Walterbos
et al. (1994) and Rix et al. (1995), is now generally assumed to mark
the acquisition of external material in a merger.  It is the outer gas
disk that `counter-rotates' (Rix et al. 1995), and the inner gas disk
may have contracted from gas that was originally at greater radii but
lost much of its angular momentum in the merger (Sil'chenko 1996).
The central region is a weak source of radio emission
($S_{1.5GHz}\,=\,100$ mJy).  HI maps have been presented by Braun et
al. (2007) and Haan et al. (2008) who found a total HI mass 
$M_{\rm HI}\,=1.8\times10^{8}\,\Msun$. The galaxy was
mapped in $\co$ and observed in $\13co$ in both the $J$=1-0 and th
$J$=2-1 transition by Casoli $\&$ Gerin (1993), further studied in HI
and in $J$=3-2 $\co$ by Braun et al. (1994).  High-resolution
millimeter array observations have been provided in the $J$=1-0 $\co$
line by Sakamoto et al. (1999, NMA, $\approx4''$), in the $J$=1-0 and
$J$=2-1 $\co$ lines by Garc\'ia-Burillo et al (2003, PdB, 1-3$''$)
and in the $J$=1-0 $\co\,\13co$ and HCN lines by Helfer et al. (2003)
and Rosolowsky $\&$ Blitz (2005).

%Table 4
\begin{table*}[]
\caption[]{Measured central $^{12}$CO and $^{13}$CO line intensities}
\begin{center}
\begin{tabular}{lrcrcrcrcrcr}
\hline
\noalign{\smallskip}
& Beam & \multicolumn{2}{l}{NGC~1068} & \multicolumn{2}{l}{NGC~2146} & \multicolumn{2}{l}{NGC~3079} & \multicolumn{2}{l}{NGC~4826} & \multicolumn{2}{l}{NGC~7469} \\
& Size$^{a}$  
& $T_{\rm mb}$ & $\int T_{\rm mb}$d$V$ 
& $T_{\rm mb}$ & $\int T_{\rm mb}$d$V$ 
& $T_{\rm mb}$ & $\int T_{\rm mb}$d$V$ 
& $T_{\rm mb}$ & $\int T_{\rm mb}$d$V$ 
& $T_{\rm mb}$ & $\int T_{\rm mb}$d$V$ \\
& ($\arcsec$)       & (mK) & ($\kkms$)  
                    & (mK) & ($\kkms$)  
		    & (mK) & ($\kkms$)  
		    & (mK) & ($\kkms$)  
		    & (mK) & ($\kkms$)  \\
\noalign{\smallskip}
\hline
\noalign{\smallskip}
$\co$ \\
$J$=1-0 & 21 & 645 & 177$\pm$21 &  770 & 196$\pm$20 &  608 & 235$\pm$28 & 405 &96$\pm$10$^{b}$ &  243 &  54$\pm$7  \\
$J$=2-1 & 12 &1080 & 266$\pm$32 & 1125 & 224$\pm$22 & 1085 & 454$\pm$54 & 425 &  82$\pm$8$^{b}$&  545 & 124$\pm$15 \\
        & 21 & 690 & 220$\pm$26 & 580  & 188$\pm$20 &  516 & 169$\pm$20 & 540 & 112$\pm$11&  203 &  50$\pm$6  \\
  & {\it 43} & --- &  99$\pm$12 & ---  &  74$\pm$7  & ---  &  94$\pm$11 & --- &  53$\pm$5 &  --- &  17$\pm$2  \\
$J$=3-2 & 14 & 630 & 166$\pm$20 & 1187 & 252$\pm$26 &  640 & 236$\pm$29 & 365 & 100$\pm$10&  197 &  93$\pm$12 \\
  & {\it 21} & --- & 114$\pm$14 & ---  & 153$\pm$17 & ---  & 138$\pm$17 & --- &  63$\pm$6 &  --- &  45$\pm$6  \\
$J$=4-3 & 11 &1060 & 290$\pm$35 &  980 & 141$\pm$15 &  700 & 254$\pm$30 & 700 & 146$\pm$15&  280 &  94$\pm$12 \\
  & {\it 14} & --- & 231$\pm$28 & ---  & 122$\pm$13 & ---  & 218$\pm$26 & --- & 139$\pm$14&  --- &  69$\pm$8  \\
  & {\it 21} & --- & 150$\pm$18 & ---  &  99$\pm$10 & ---  & 144$\pm$17 & --- &  83$\pm$8 &  --- &  ---       \\
\noalign{\smallskip}
\hline
\noalign{\smallskip}
$\13co$ \\
$J$=1-0 & 21 &  51 &15.2$\pm$1.8&   45 & 12.9$\pm$1.3&  36 & 14.7$\pm$1.8 & 43 & 11.7$\pm$1.2$^{a}$&  15 & 3.3$\pm$0.4 \\
$J$=2-1 & 12 &  56 &15.0$\pm$1.8&  104 & 19.0$\pm$2.1&  82 & 29.3$\pm$3.6 & 62 & 13.2$\pm$1.6$^{a}$&  33 & 5.7$\pm$0.9 \\
        & 21 &  46 &17.1$\pm$2.0&  106 & 16.8$\pm$1.8&  26 & 10.8$\pm$1.3 & 65 & 14.9$\pm$1.7      &  25 & 4.2$\pm$0.7 \\
$J$=3-2 & 14 &  43 &10.7$\pm$1.3&   86 & 18.3$\pm$1.9&  60 & 26.0$\pm$3.1 & 42 &  9.3$\pm$1.1      &  25 & 3.5$\pm$0.7 \\
\noalign{\smallskip}
\hline
\end{tabular}
\end{center}
\label{galintensity}
Notes:$^{a}$ beams not directly observed but obtained from convolved
maps (see Section\,3.2) are in italics; $^{b}$ measured position is
10$''$ off nucleus.
\end{table*}

\subsection{NGC~7469 = Arp~298}

This is the most distant galaxy in our sample. It is a classical
Seyfert 1 galaxy, and it is well-studied at various wavelengths. It
appears to be interacting with its companion IC~5283, $1.3'$ to the
NNE.  It should be noted that various distances quoted in the
literature are about twice the distance assumed in this paper.
High-resolution VLA radio maps have been presented by e,g. Ulvestad et
al. (1981); Condon et al. (1982, 1990) and Wilson et al. (1991). The
radio emission (total flux density $S_{1.5GHz}$ = 160 mJy) extends
over the central $\approx\,8''$ (1150 pc) and arises mostly from a
circumnuclear starburst `ring' (Wilson et al. 1986, 1991) surrounding
a compact nuclear radio source (flux density S$_{4.9GHz}$ = 21
mJy). The nucleus is a variable X-ray source (Barr, 1986; Fabbiano et
al. 1992). The circumnuclear starburst dominates the energy budget as
it represents two thirds of the bolometric luminosity of the entire
galaxy (Genzel et al 1995) and is on the order of $!0^{11}$ L$_{\odot}$. The
galaxy is a powerful object at infrared wavelengths and was mapped
with millimeter arrays in the $J$=1-0 $\co$ line by Sanders et
al. (1988, OVRO, $6''$) and Meixner et al. (1990, Hat Creek, $2''$),
as well as more recently in the $J$=2-1 $\co$ and $J$=1-0 HCN lines by
Davies et al. (2004, PdB, $0.7''$ and $2''$ respectively).
Significant detail is, however, only revealed by the very
high-resolution $J$=2-1 $\co$ map.

\section{Observations and results}

\subsection{Observations}

The observations described in this paper were carried out with the 15m
James Clerk Maxwell Telescope (JCMT) on Mauna Kea (Hawaii)
\footnote{The James Clerk Maxwell Telescope is operated on a joint
basis between the United Kingdom Particle Physics and Astrophysics
Council (PPARC), the Netherlands Organisation for Scientific Research
(NWO) and the National Research Council of Canada (NRC).} and with the 
IRAM 30m telescope on Pico Veleta (Spain).
\footnote{The IRAM 30m telescope is supported by INSU/CNRS (France), 
MPG (Germany), and IGN (Spain).}

\subsubsection{JCMT Observations}  

At the epoch of the mapping observations (1991-2001) the absolute
pointing of the telescope was good to about $3''$ r.m.s. as provided
by pointing observations with the JCMT submillimeter bolometer.  The
spectra were calibrated in units of antenna temperature $T_{\rm
  A}^{*}$, correcting for sideband gains, atmospheric emission in both
sidebands and telescope efficiency.  Calibration was regularly checked
by observation of a standard line source.  Further observational
details are given in Tables\,\ref{gal12colog} and \ref{gal13colog}.
Most of the observations were carried out with the now defunct
receivers A2, B3i and C2. Observations in 2001 were obtained with the
current receivers B3 (330/345 GHz) and W/C (461 GHz). Full details on
these receivers can be found at the JCMT website
(http://docs.jach.hawaii.edu/JCMT/HET/GUIDE/).  Up to 1993, we used a
2048-channel AOS backend covering a band of 500 MHz ($650\kms$ at 230
GHz). After that year, the DAS digital autocorrelator system was used
in bands of 500 and 750 MHz. Integration times (on+off) given in
Tables\,\ref{gal12colog} are typical values appropriate to the maps.
We subtracted second order baselines from the profiles.  All spectra
were scaled to a main-beam brightness temperature, $T_{\rm mb}$ =
$T_{\rm A}^{*}$/$\eta _{\rm mb}$; values for the values of $\eta _{\rm
  mb}$ used are given in Tables\,\ref{gal12colog}\,\ref{gal13colog}.

All maps were made in rectangular grids (parameters given in
Table\,\ref{gal12colog}), rotated to parallel the major axis of the galaxy
mapped.  The maps shown in this paper all have been rotated back to a
regular grid in right ascension and declination.  Some additional
remarks are in order.  In the case of NGC~1068, we combined our data
as listed in Tabel\,\ref{gal12colog} with those obtained by Papadopoulos
$\&$ Seaquist (1999), taken from the JCMT archives, and re-reduced. The
CO maps presented in this paper are based on both datasets.  The
velocity widths of the lines from the edge-on galaxy NGC~3079 were
similar to the bandwidth of the JCMT backends used. For this reason,
we observed the lines at three slightly different velocity
settings. The resulting spectra were combined and concatenated to
yield the final spectra.

\subsubsection{IRAM observations}

Observations of both $\co$ and $\13co$ in the $J$=1--0 and $J$=2--1
transitions at resolutions of $21''$ and $12''$ respectively were
obtained towards the centers of all galaxies with the IRAM 30m
telescope in the period 2005-2007.  We used the IRAM facility
low-noise receivers A and B in both the 3mm and the 1mm bands. For
backends we used the 1MHz and 4MHz filter banks, as well as the VESPA
correlator.  Initially, all four transitions were observed
simultaneously, therefore with identical pointings. After this, the
signal-to-noise ratio of the $\13co$ profiles was increased by
additional simultaneous observation of the two transitions, where we
took care to ensure that the pointing was the same as for the earlier
observations.  Calibration and reduction procedures were standard and
similar to those described in the previous section.  The assumed $\eta
_{\rm mb}$ values are likewise given in Tables\,\ref{gal12colog} and
\ref{gal13colog}. The IRAM observations were especially important not
only because they provide measurements of the 3mm $J$=1--0 transitions
not possible with the JCMT. The IRAM aperture is twice that of the
JCMT, so that the combination of the two datasets provides
$J$=2-1/$J$=1-0 and $J$=3-2/$J$=2-1 ratios in closely-matched beams,

\subsection{Results}

Central line profiles in all observed transitions are shown in
Figs.\,\ref{n1068COmaps} through \ref{n7469COmaps}.  We mapped the
distribution of the molecular gas in the galaxy centers in the
$J$=2--1, $J$=3--2, and $J$=4--3 transitions, typically over an area
of about one arcminute across.  These maps are also shown in the
Figures, as are the major axis position velocity diagrams of the
central CO emission.

Measured line intensities towards the center of each galaxy are listed
in Table\,\ref{galintensity}.  This Table contains entries both at
full resolution, and at resolutions corresponding to those of
transitions observed in larger beams.  These entries were extracted
from maps convolved to the resolution quoted. The convolution was
carried out with the map interpolation function of the SPECX data
reduction package\footnote{
http://docs.jach.hawaii.edu/JCMT/cs/005/11/html/node129.html}. This
function interpolates, where necessary, the map intensities by using
adjacent pixels out to a preset distance (usually 2 beams) and
convolves the observed and interpolated data points with a
two-dimensional Gaussian function (`beam') of predetermined halfwidth.
This halfwidth is chosen such that the actual observing beam convolved
with the 2D-Gaussian would yield a beam corresponding to the desired
resolution.  Velocities in a position-velocity map can also be
`smoothed' by applying a convolving Gaussian function in a similar
manner.

Finally, we have determined the line intensity ratios of the observed
transitions at the center of the galaxies observed.  These line ratios
have been corrected for small offsets between the various datasets.
They have been determined independently from the intensities given in
Table\,\ref{galintensity}. Specifically: (a) we used data taken at the
same observing run wherever possible, (b) we determined ratios at
various resolutions convolving the maps to the relevant beam widths,
(c) we verified ratios by comparing profile shapes in addition to
integrated intensities, and (d) the line ratios have been corrected
for small offsets between the various datasets. In Table 5, whenever
data were insufficient for convolution, or altogether lacking
(e.g. [CI]), no entry is provided.

%Table 5  Line Ratios
\begin{table*}
\begin{center}
\caption[]{Adopted velocity-integrated line ratios in galaxy centers}
\begin{tabular}{lcccccc}
\hline
\noalign{\smallskip}
Transitions    & Beam   & NGC~1068      & NGC~2146      & NGC~3079      & NGC~4826      & NGC~7469     \\
               & ($'')$ &               &               &               &               & \\
\noalign{\smallskip}
\hline
\noalign{\smallskip}
$\co$ \\
(1--0)/(2--1)  & 21     & 0.97$\pm$0.18 & 1.04$\pm$0.17 & 1.39$\pm$0.27 & 1.02$\pm$0.17 & 1.08$\pm$0.16 \\
(2--1)/(2--1)  & 12/21  & 1.21$\pm$0.24 & 1.19$\pm$0.17 & 2.69$\pm$0.54 & 1.08$\pm$0.22 & 2.47$\pm$0.38 \\
(3--2)/(2--1)  & 21     & 0.52$\pm$0.11 & 0.82$\pm$0.12 & 0.72$\pm$0.14 & 0.60$\pm$0.09 & 0.86$\pm$0.13 \\
(3--2)/(2--1)  & 12     & 0.71$\pm$0.14 & 0.99$\pm$0.15 & 0.61$\pm$0.12 & 0.84$\pm$0.18 & 0.87$\pm$0.13 \\
(4--3)/(2--1)  & 21     & 0.68$\pm$0.13 & 0.65$\pm$0.16 & 0.76$\pm$0.15 & 0.74$\pm$0.12 &  ---          \\
(4--3)/(2--1)  & 12     & 1.00$\pm$0.20 & 0.63$\pm$0.11 & 0.56$\pm$0.11 &     ---       & 0.66$\pm$0.11 \\
\noalign{\smallskip}
$\co$/$\13co$ \\
(1--0)         & 21     & 11.9$\pm$1.8  & 15.2$\pm$2.0  & 16.2$\pm$2.4  &  8.2$\pm$1.2  & 16.2$\pm$2.4  \\
(2--1)         & 21     & 11.4$\pm$1.7  & 11.2$\pm$1.7  & 13.5$\pm$2.0  &  7.2$\pm$1.2  & 16.4$\pm$1.7  \\
(2--1)         & 12     & 15.8$\pm$2.4  & 11.8$\pm$1.5  & 15.2$\pm$2.2  &  6.2$\pm$0.9  & 20.5$\pm$3.0  \\
(3--2)         & 14     & 15.3$\pm$2.2  & 13.3$\pm$2.0  &  9.1$\pm$1.4  & 10.8$\pm$2.6  & 15.3$\pm$2.3  \\
(3--2)         & 21     & 13.3$\pm$2.0  &      ---      &      ---      &      ---      &      ---      \\
\noalign{\smallskip}
[CI]/CO(2--1))   & 21     & 0.33$\pm$0.07 &      ---      & 0.57$\pm$0.11 & 0.58$\pm$0.23 &      ---      \\
\noalign{\smallskip}
\hline
\end{tabular}
\end{center}
\label{galratio}
\end{table*}

%Table 6
\begin{table*}
\caption[]{Radiative transfer model parameters}
\begin{center}
\begin{tabular}{lccccccccccccc}
\hline
\noalign{\smallskip} 
No. & \multicolumn{3}{c}{Component 1} & \multicolumn{3}{c}{Component 2} & Ratio &\multicolumn{6}{c}{Model Ratios}\\
     & Kin.       & Gas	& Gradient & Kin.   & Gas      & Gradient    & Comp. &\multicolumn{3}{c}{$\co$} &\multicolumn{3}{c}{$\co/\13co$}\\
     & Temp.      & Dens.    & $\textstyle{N(CO)\over dV}$ & Temp.  & Dens.    & $\textstyle{N(CO)\over dV}$ & 1:2   &$\textstyle{(1-0)\over(2-1)}$&$\textstyle{(3-2)\over(2-1}$&$\textstyle{(4-3)\over(2-1)}$ & (1-0) & (2-1) & (3-2) \\
     & $T_{\rm k}$  & $n(\h2$) & 	  & $T_{\rm k}$ & $n(\h2$)    & $N$(CO)/d$V$&\\
     & (K)     	    & ($\cc$)    & ($\textstyle{\cm2\over\kms}$) & (K) 	& ($\cc$) &($\textstyle{\cm2\over\kms}$) &\\
\noalign{\smallskip}
\hline
\noalign{\smallskip}
\multicolumn{14}{l}{NGC~1068} \\
 1 &  60 &   1000 &  3.0$\times10^{17}$ &  30 &   3000 & 0.06$\times10^{17}$ & 1:9 &1.08&0.62&0.32&10.9&10.9&13.2\\
 2 &  30 &   3000 &  1.0$\times10^{17}$ &  30 &   3000 & 0.06$\times10^{17}$ & 1:4 &0.99&0.62&0.28&13.2&10.6&12.2\\
\noalign{\smallskip}
\multicolumn{14}{l}{NGC~2146} \\
 3 & 125 &   1000 &  1.0$\times10^{17}$ &  30 & 100000 &  0.3$\times10^{17}$ &11:9 &1.09&0.86&0.64&15.5&11.3&13.3\\
 4 & 150 &   1000 &  1.0$\times10^{17}$ & 125 &  30000 &  1.0$\times10^{17}$ & 6:1 &1.06&0.83&0.60&15.3&11.5&13.3\\
\noalign{\smallskip}
\multicolumn{14}{l}{NGC~3079} \\
 5 & 150 &    100 &  0.3$\times10^{17}$ &  30 & 100000 &  0.6$\times10^{17}$ & 6:1 &1.28&0.69&0.51&16.3&13.3& 9.8\\
 6 & 150 &   1000 &  1.0$\times10^{17}$ &  20 &  10000 &  0.3$\times10^{17}$ & 1:4 &1.14&0.72&0.40&16.5&14.2&10.2\\
\noalign{\smallskip}
\multicolumn{14}{l}{NGC~4826}\\ 
 7 &  60 &   3000 &  1.0$\times10^{17}$ &  10 &   1000 &  1.0$\times10^{17}$ & 1:2 &1.19&0.70&0.43& 8.0& 7.6&10.3\\
\noalign{\smallskip}
\multicolumn{14}{l}{NGC~7469 } \\
 8 &  60 & 100000 &  0.6$\times10^{17}$ &  30 &    500 &  0.3$\times10^{17}$ & 1:6 &1.06&0.80&0.65&15.8&18.5&16.4\\
 9 &  30 & 100000 &  0.3$\times10^{17}$ &  10 &    100 &  0.6$\times10^{17}$ & 1:6 &1.09&0.79&0.66&17.2&18.1&13.9\\
\noalign{\smallskip}
\hline
\end{tabular}
\end{center}
\label{galmodel}
\end{table*}

%Table 7
\begin{table*}
\caption[]{Beam-averaged model parameters}
\begin{center}
\begin{tabular}{lccccccccc}
\hline
\noalign{\smallskip} 
Galaxy & No. & \multicolumn{3}{c}{Beam-Averaged Column Densities} & Outer  & Central  & Face-on      & Relative \\
       & &         &        &         & Radius & $\h2$ Mass & Mass Density & Mass \\
       & & $N$(CO) & $N$(C) & $N(\h2)$ & $R$ & $M_{\h2}$ & $\sigma(\h2)$ & Components 1:2 \\
   & & \multicolumn{2}{c}{($10^{18} \cm2$)} & ($10^{21} \cm2$) & (kpc) & ($10^{8} \Msun$) & ($\Msun$/pc$^{-2}$) & \\
\noalign{\smallskip}
\hline
\noalign{\smallskip}
NGC~1068 & 1/2 & 0.60& 0.25& 1.6 & 1.0 & 1.2 &  21 & 0.20:0.80 \\
NGC~2146 & 3   & 0.50& 0.40& 1.6 & 1.5 & 1.4 &  18 & 0.75:0.25 \\
         & 4   & 0.70& 0.55& 2.3 & 1.5 & 2.0 &  26 & 0.80:0.20 \\
NGC~3079 & 5   & 1.20& 0.50& 3.0 & 0.9 & 1.8 &  19 & 0.80:0.20 \\
         & 6   & 1.50& 0.50& 3.6 & 0.9 & 2.1 &  22 & 0.40:0.60 \\
NGC~4826 & 7   & 0.85& 0.45& 2.4 & 0.3 & 0.3 &  23 & 0.30:0.70 \\
NGC~7469 & 8/9 & 0.20& 0.45& 1.2 & 1.2 & 2.5 &  15 & 0.10:0.90 \\
\noalign{\smallskip}
\hline
\end{tabular}
\end{center}
\label{galresult}
\end{table*}

\section{Analysis}

\subsection{Radiative transfer modeling of CO}

We have modeled the observed $\co$ and $\13co$ line intensities and
ratios with the large velocity gradient (LVG) radiative transfer
models described by Jansen (1995) and Jansen et al. (1994) -- but see
also Hogerheijde $\&$ van der Tak (2000) and the web page
http://www.strw.leidenuniv.nl/$\,\tilde{}\,$michiel/ratran/.  These
codes provide model line intensities as a function of three input
parameters per molecular gas component: gas kinetic temperature
$T_{\rm k}$, molecular hydrogen density $n(H_{2})$ and the CO column
density per unit velocity $N({\rm CO})$/d$V$.  By comparing model to
observed line {\it ratios}, we may identify the physical parameters
best describing the actual conditions at the observed positions.
Beam-averaged properties are determined by comparing observed and
model intensities. In principle, with seven measured line intensities
of two isotopes, properties of a single gas component are
overdetermined as only five independent observables are required.  In
practice, this is mitigated by degeneracies such as occur for
$^{12}$CO.  In any case, we find that fits based on a single-component
gas are almost always incapable of fully matching the data.  However,
we usually obtain good fits based on {\it two} gas components.  The
solutions for a two-component gas are slightly underdetermined but we
can succesfully compensate for this by introducing additional
constraints as described below.  The physical gas almost certainly has
a much wider range of temperatures and densities.  However, with the
present data two components is the maximum that can be considered, as
adding even one more component increases the number of free parameters
to the point where solutions are physically meaningless.  Thus, our
two-component model gas is only an approximation of reality but it is
decidedly superior to one-component model solutions.  As long as
significant fractions of the total molecular gas mass are limited to
similar segments of parameter space, they also provide a reasonably
realistic model for the actual state of affairs.  Specifically, our
analysis is less sensitive to the occurrence of gas at very high
densities and temperatures, but such gas is unlikely to contribute
singificantly to the total mass.

In order to reduce the number of free parameters, we assumed identical
CO isotopical abundances for both gas components.  Furthermore, in a
small number of starburst galaxy centers (NGC~253, NGC~4945, M~82,
IC~342, He~2-10), values of $40\pm10$ have been suggested for the
isotopical abundance [$^{12}$CO]/[$^{13}$CO] (Mauersberger $\&$ Henkel
1993; Henkel et al. 1993, 1994, 1998; Bayet et al. 2004), somewhat
higher than the characteristic value of 20--25 for the Milky Way
nuclear region (Wilson \& Rood 1994).  We therefore have adopted an
abundance value of [$^{12}$CO]/[$^{13}$CO] = 40 in our models.  We
identified acceptable fits by searching a grid of model parameter
combinations (10 K $\leq T_{\rm k} \leq $ 150 K, $10^{2} \cc \leq
n(\h2) \leq 10^{5} \cc$, $6 \times 10^{15} \cm2 \leq N(CO)/dV \leq 3
\times 10^{18} \cm2$) for model line ratios matching the observed set,
with the relative contribution of the two components as a free
parameter.  Solutions obtained in this way are not unique, but rather
define a limited range of values in distinct regions of parameter
space.  For instance, variations in input parameters may to some
extent compensate one another, producing identical line ratios for
somewhat different combinations of input parameters.  Among all
possible solution sets, we have rejected those in which the denser gas
component is also much hotter than the more tenuous component, because
we consider the large pressure imbalances implied by such solutions
physically implausible, certainly on the kiloparsec scales observed.

The results of our model fitting procedure are summarized in
Table\,\ref{galmodel}. 

\subsection{Beam-averaged molecular gas properties}

The chemical models presented by van Dishoeck $\&$ Black (1988) show a
strong dependence of the $N({\rm C})/N({\rm CO})$ column density ratio
on the total carbon (C+CO) and molecular hydrogen column densities.
If line intensities of all three species, CO, [CI] and [CII] are
known, minimal assumptions suffice to deduce total carbon and hydrogen
column densities and masses.

If C$^{\circ}$ and C$^{+}$ have not been observed, we can still relate
the radiative transfer model results to molecular hydrogen column
densities by taking the [C]/[H] gas-phase abundance ratio as an
additional constraint.  We must then find the molecular hydrogen
column density that satisfies both the total carbon $N_{\rm
  C}\,=\,N(C)\,+\,N(CO)$ column-density {\it and} the carbon ratio
$f_{\rm C}\,=\,N(C)/N(CO)$ commensurate with the above-mentioned
models and [C]/[H] abundance.  We intend to compare the results for
various galaxies, several of which are lacking either [CI] or [CII]
measurements, or both.  Thus, for the sake of consistency, we use this
second method for all galaxies, whether or not [CI] or [CII] have been
detected.

Oxygen abundances and gradients of many nearby galaxies have been
compiled by Vila-Costas $\&$ Edmunds (1992), Zaritsky et al. (1994),
and Pilyugin, V\'ilchez $\&$ Contini (2004).  Extrapolated to zero
radius, the first two compilations imply nuclear metalicities many
times solar.  However, Pilyugin et al.'s more recent compilation
yields nuclear metalicities on average only 1.5 times solar.  These
are nevertheless quite uncertain, because the galaxy disk values and
gradients were derived from HII regions at various distances to the
center, and it is not very clear whether such results can be
extrapolated to the rather different galactic center conditions.
Another practical complication is that of our sample, only NGC~1068 is
included in these compilations, although we have an estimate for
NGC~7469 from X-ray observations (Blustin et al. 2007).  In the
following, we have conservatively assumed a metalicity twice solar to
apply to the galaxy central regions.  The results published by Garnett
et al. (1999) suggest very similar carbon and oxygen abundances at
these metalicities, so that we adopt [C]/[H]
$\approx\,1.0\,\times\,10^{-3}$.  As a significant fraction of carbon
is tied up in dust particles and thus unavailable in the gas-phase, we
have also adopted a fractional correction factor $\delta_{\rm c}$ =
0.27 (see for instance van Dishoeck $\&$ Black 1988), so that $N_{\rm
  H}/N_{\rm C}$ = [2$N(\h2) + N$(HI)]/[$N$(CO) + $N$(CII) + $N$(CI)] =
3700, uncertain by about a factor of two.  In Table\,\ref{galresult}
we present beam-averaged column densities for CO and C (=C$^{\rm
  o}$+C$^{+}$), as well as $\h2$ derived under the assumptions just
discussed.  As the observed peak CO intensities are significantly
below the model peak intensities, only a small fraction of the (large)
beam surface area can be filled with emitting material.

Although the analysis in terms of two gas components is superior to
that assuming a single component, it is still not fully realistic.
For instance, the assumption of e.g. {\it identical beam filling
  factors} for the various species ($\co, \13co$, C$^{\circ}$, and
C$^{+}$) is not a priori plausible.  Fortunately, these assumptions
are useful but not critical in the determination of beam-averaged
parameters.  If, by way of example, we assume a smaller beam filling
factor, the model cloud intensity increases. This generally implies a
higher model column-density which, however, is more strongly diluted
by the beam.  The beam-averaged column-density is modified only by the
degree of non-linearity in the response of the model parameters to a
change in filling factor, {\it not} by the magnitude of that change. A
similar state of affairs exists with respect to {\it model degeneracy},
i.e different model parameter combinations yielding (almost) identical
line ratios.  Although, for instance, the model fractions of hot or
dense gas may differ, the final derived {\it beam-averaged} molecular
hydrogen densities are practically the same.  In fact, the assumed
gas-phase carbon abundance dominates the results.  The derived
beam-averaged column-densities (and projected mass densities) roughly
scale inversely proportional to the square root of the abundance
assumed: $N_{av}\,\propto\,{[C]/[H]_{gas}}^{-0.5}$. As our estimate of
the central elemental carbon abundance, and our estimate of the carbon
depletion factor are unlikely to be wrong by more than a factor of two, the
resulting uncertainty in the final values listed in Table\,\ref{galresult}
should be no more than $50\%$.

\section{Discussion}

\subsection{NGC~1068}

{\it The extended inner star-forming zone} has line ratios that do not
allow for much variation of the gas properties.  Given the size
(deconvolved dimensions of $27\times24''$) of the region, its general
appearance, and the relatively slight contribution of the compact
circumnuclear disk, the emission in the lower CO transitions refers to
the extended emission associated with the star-forming spiral arms and
disk (see, for instance, Wynn-Williams, Becklin $\&$ Scoville 1985 or
Le Floc'h et al. 2001). The major axis-velocity map shown in in
Fig.\,\ref{n1068COmaps} illustrates the dominance of the spiral arms
in the $J$=3-2 $\co$ distribution in the shape of the bright maxima at
the extreme velocities, with much less bright CO emission closer to
the nucleus. The central minimum is not pronounced, however, because
it is filled in by emission from the unresolved circumnuclear source
(see also Fig. 6 in Helfer $\&$ Blitz 1995).  We did not use the
$J$=4-3 results in our analysis.  The models predict a $J$=4-3/$J$=2-1
line ratio of about 0.3, only half of that actually observed
(Table\,\ref{galratio}).  However, much of the emission in this
transition arises in the compact circumnuclear gas, rather than in the
extended source (see Fig.\,\ref{n1068comp}.  We find a relatively
cool molecular gas ($T_{\rm kin} \approx 30$ K) with densities ranging
over ($n_{\h2}\,=\,10^{3}$-$10^{4}\,\cc$).  The inner star-forming zone
shows up quite clearly in the high-resolution HI map by Brinks et
al. (1997), reaching peak column densities
$N(HI)\,\approx\,3\times10^{21}\,\cm2$ in $8''$ beams. Such $\h2$ and
HI values are typical values for photon-dominated regions (PDRs)
associated with star formation.

Most, but not all (60$\%$) carbon must be locked up
in CO.  The overall filling factor of the molecular gas is $7\%$. The
molecular gas mass mapped over the central $2R$ = 2 kpc is slightly
more than $M_{\h2}\,=\,10^{8}\,\Msun$, i.e a bit less than one
per cent of the dynamical mass $M\,\approx\,1.5\,\times\,10^{10}$
implied by the major-axis-velocity map in Fig\,\ref{n1068COmaps}.

%Figure6: More NGC1068 Maps
\begin{figure}[]
\unitlength1cm
\begin{minipage}[t]{8.5cm}
\resizebox{4cm}{!}{\rotatebox{270}{\includegraphics*{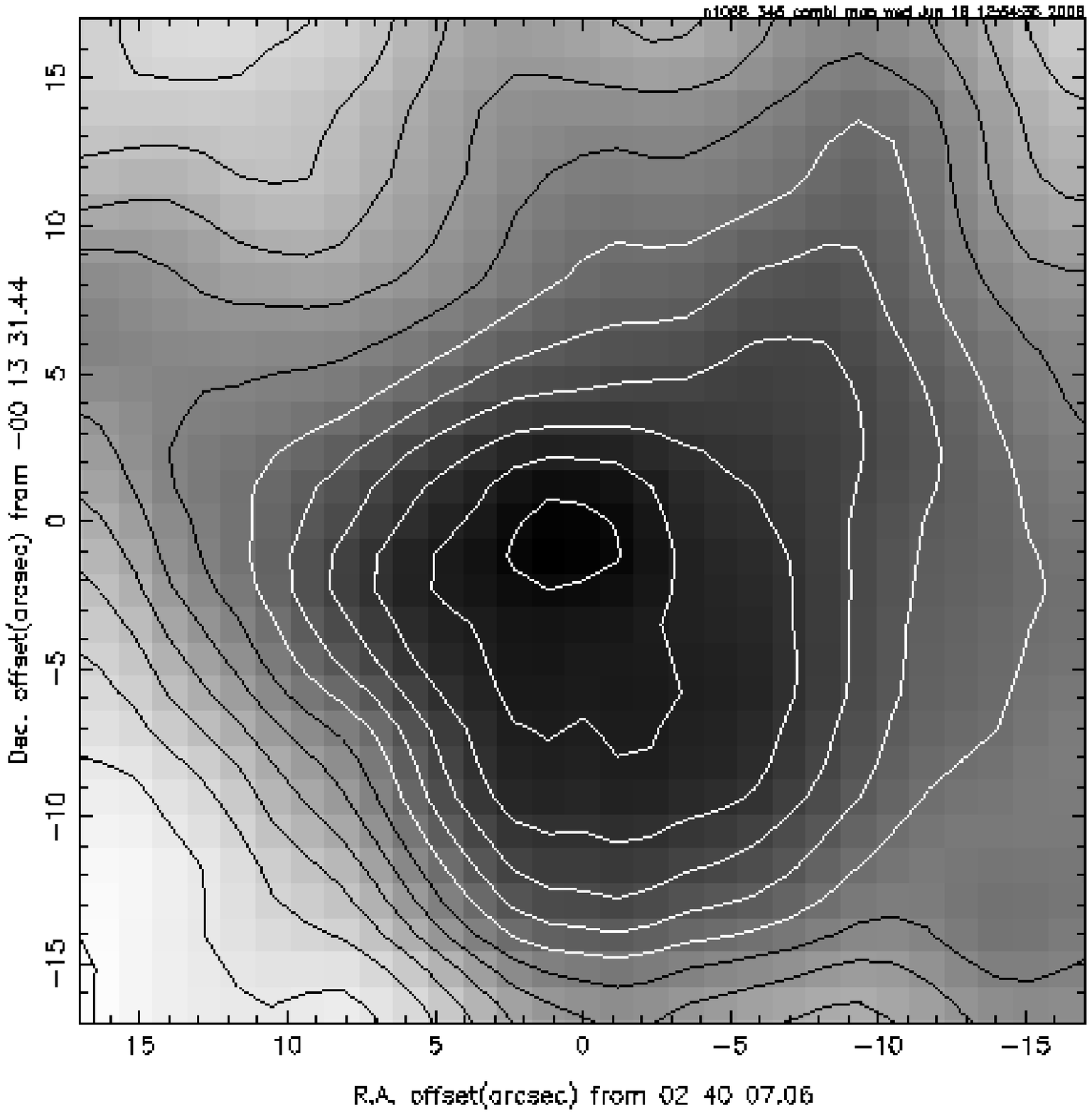}}}
\hfill
\resizebox{4cm}{!}{\rotatebox{270}{\includegraphics*{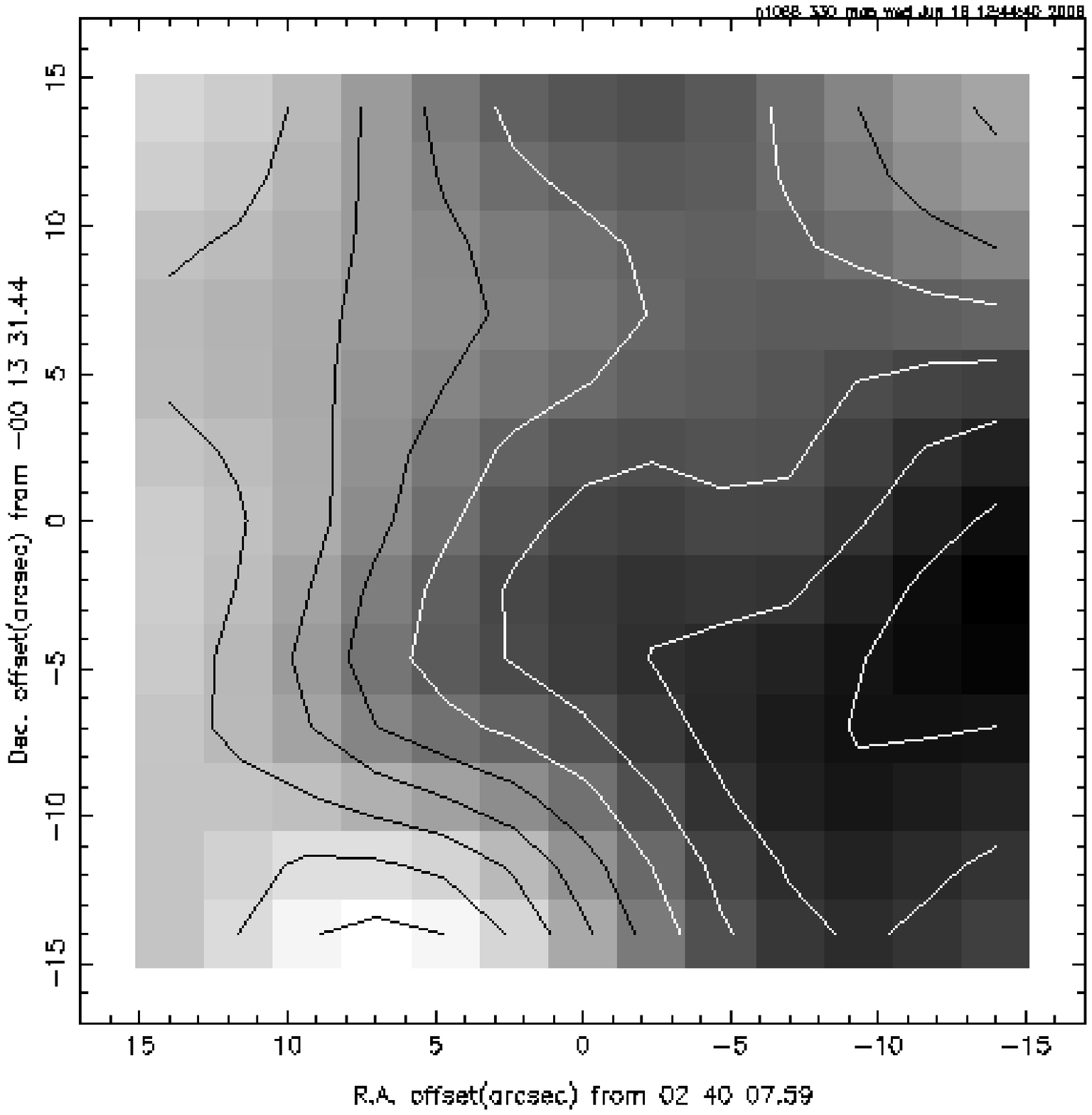}}}
\end{minipage}

\vspace{0.2cm}
\begin{minipage}[t]{8.5cm}
\resizebox{4cm}{!}{\rotatebox{270}{\includegraphics*{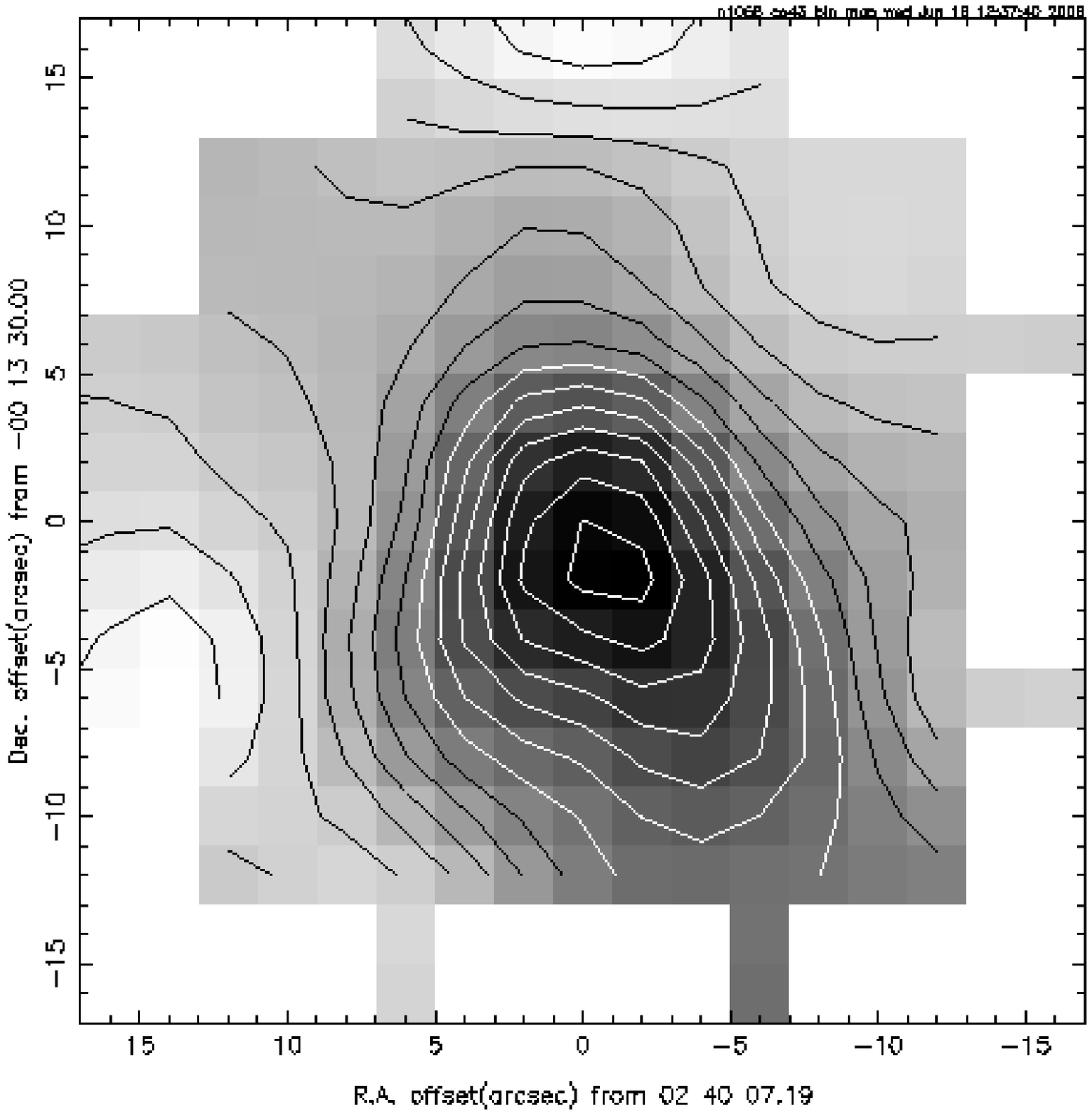}}}
\hfill
\resizebox{4cm}{!}{\rotatebox{270}{\includegraphics*{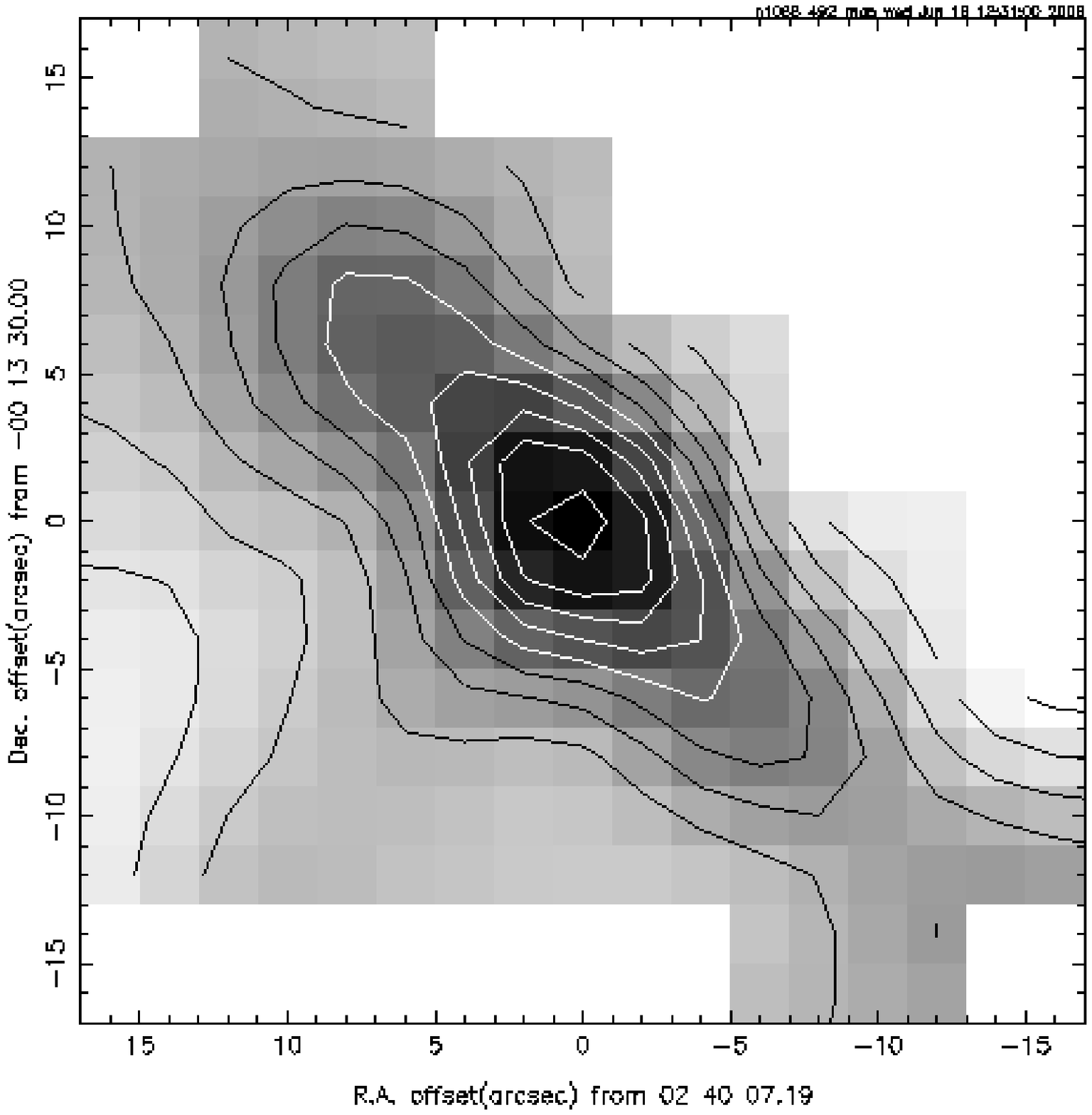}}}
\end{minipage}
\caption[] {Central $30''$ of NGC~1068.  Top left $J=3-2$ $\co$ with
  contour step of 10 $\kkms$, top right $J=3-2$ $\13co$ with contour
  step of 1 $\kkms$, bottom left $J=4-3$ $\co$ with contour step of 20
  $\kkms$, bottom right [CI] with contour step of 10 $\kkms$. Note
  dominance of the unresolved nuclear source in the latter two maps}
\label{n1068comp}
\end{figure}

{\it The compact circumnuclear gas} is reasonably well resolved by the
10$''$-14$''$ beam data, but the lack of $\13co$ intensities severely
hamper attempts to model it. The ratio $\co/\13co\,>\,6$ implied by
the BIMA observations by Helfer $\&$ Blitz (1995) does not present a
significant constraint.  Nevertheless, the models do imply that a
significant fraction (10-50$\%$) of this circumnuclear gas must have a
density $n_{\h2}\,\approx\,10^{5}\,\cc$ and a temperature of 30-60 K.
This is consistent with the values estimated by Jackson et al. (1993),
Tacconi et al. (1994), Helfer $\&$ Blitz (1995), Papadopoulos $\&$
Seaquist (1999), and Usero et al. (2004). The remainder of the gas
must be rather tenuous ($n_{\h2}\,=\,100-1000\,\cc$) at an undetermined
temperature.  The emission from this gas is particularly prominent in
the $J$=4-3 $\co$ and [CI] maps shown in Fig.\,\ref{n1068comp}, and
the maps by Jackson et al. (1993) and Helfer $\&$ Blitz (1995) show a
similar prominent role for $J$=1-0 HCN.  The lack of $\13co$ emission
precisely at the nucleus where the [CI] emission is strongest, implies
that the actual [CI]/$\13co$(2-1) ratio is significantly higher than
the beam-confused value given by Israel $\&$ Baas (2002) -- already one
of the highest in their galaxy sample.  Inside the star-forming spiral
arms, and thus also towards the nucleus, there is very little HI
($N(HI)\,\leq\,2\times10^{20}\,\cm2$ {\it cf} Brinks et al. 1997).  Much
of the circumnuclear gas thus is dense and warm, and appears to be
excited by the AGN rather than by luminous stars ({\it cf} Lepp $\&$ Dalgarno
1996; Usero et al. 2004, Meijerink et al. 2006, 2007).  Strong neutral
carbon emission is characteristic for X-ray excitation (Meijerink et
al. 2007).

{\it The extended bar.} Particularly interesting is the structure
exhibited by [CI].  The unresolved intense source, coinciding with the
nucleus, is accompanied by tongues of [CI] emission stretching out to
the northeast and southwest in a position angle of about $45^{\circ}$.
The two extensions are real, and not due to instrumental effects, as
the [CI] profiles in the map accurately represent the rapidly changing
velocity field.  This extended [CI] emission is in the same position
angle as the northeastern radio jet (Wilson $\&$ Ulvestad 1987), and
coincides with the northeastern near-infrared [FeII] line and X-ray
continuum emission extensions (Blietz et al.1994; Young et al. 2001).
However, the carbon emission in our map also extends in the {\it
  opposite} direction and thus more closely corresponds to the
distribution of the near-infrared K-band continuum emission (Rotaciuc
et al. 1991; Davies, Sugai $\&$ Ward 1998), the near-infrared {\it
  polarized} emission (Lumsden et al. 1999), and the $J$=1-0 $\co$
emission in the fully-UV-sampled array map by Helfer $\&$ Blitz
(1995).  Interestingly, the southwestern molecular gas extension
clearly seen in their map is absent in the $\co$ maps by Schinnerer et
al. (2000) and Usero et al. (2004) that lack the short-spacing
information.  There is no HI {\it emission} corresponding to the [CI]
emission ({\it cf} Brinks et al. 1997), although some HI {\it absorption}
($N(HI)\,=1-2\times10^{21}\,\cm2$ in a $\approx2''$ beam) is seen 
against the continuum of the southwestern bar. The nature of the extended
carbon emission is not clear.  Its presence at directions devoid of
X-ray or [FeII] emission suggests that the observed strong [CI]
emission does not originate in an X-ray-excited dense gas.  Rather it
appears that this emission arises from (photo)dissociated gas
streaming along the stellar and molecular bar discussed by Tacconi et
al. (1994) and Helfer$\&$ Blitz (1995).  Its nature would thus be
different from that of the neutral carbon in the circumnuclear source.

\subsection{NGC~2146}

%Figure 7: More NGC2146 Maps
\begin{figure}[]
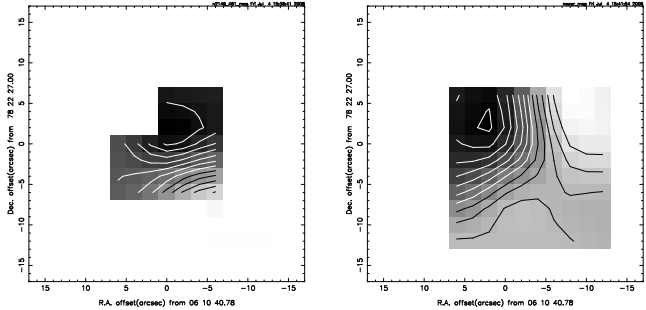

\unitlength1cm
\begin{minipage}[t]{8.5cm}
\resizebox{4cm}{!}{\rotatebox{270}{\includegraphics*{n2146_461_10kkms.ps}}}
\hfill
\resizebox{4cm}{!}{\rotatebox{270}{\includegraphics*{n2146_690_10kkms.ps}}}
\end{minipage}
\caption[] {Central $30''$ of NGC~2146.  On the left $J$=4-3 $\co$,
  and on the right $J$=6-5 $\co$. Both maps have contour steps of
  10 $\kkms$.  The relatively high intensity of $J$=6-5 $\co$
  confirms that much of the molecular gas in NGC~2146 is at elevated
  temperatures.}
\label{n2146comp}
\end{figure}

Our maps show bright CO extending over $37''$ ($J$=3-2 size)
coincident with the bright central starburst region.  The deconvolved
extent of $28''$ (2.4 kpc) is identical to that found by Dumke et
al. (2001) with a larger beam, and indeed to the extent of the
resolved $\co$ and dust emission observed at much higher resolution
with the various millimeter arrays (Young et al. 1988b; Jackson $\&$
Ho 1988: Greve et al. 2006).  Thus, there appears to be very little
molecular gas beyond radii of a kiloparsec. The CO extent also
corresponds quite closely to that of the bright radio emission in the
VLA maps by Kronberg $\&$ Biermann (1981) and Zhao et al (1996); the
array of discrete radio sources mapped at very high resolution
($0.4''$) by Tarchi et al. (2000) covers a full span of $25''$.  The
appearance of the high-resolution CO maps, the unmistakable double
structure in maps of peak rather than integrated brightness
temperature (not shown) and the shape of the CO distribution in the
major axis-velocity diagram (Fig.\,\ref{n2146COmaps}), exhibiting
minimum intensity at the galaxy center, establish that the observed CO
does not homogeneously fill the central volume.  Rather, it must be
distributed in a warped ring (Greve et al. 2006), or much more likely,
in CO-enhanced spiral arms seen mostly edge-on.  The central few
hundred parsec should be largely free of gas.  The {\it ionized} gas
measured by Zhao et al. (1996) appears to cover the full range of
velocities seen in Fig.\,\ref{n2146COmaps} {\it within} the central
$5''$.  Tarchi et al. (2002) found two water kilomasers marking
star-forming regions. One is located in the northwestern CO emitting
region, $7$ from the center with $V_{LSR}\,\approx 825\,\kms$, and the
other projected very close to the nucleus with
$V_{LSR}\,=\,1010\,\kms$, also within the CO contours in
Fig\,\ref{n2146COmaps}-bottom.

Our LVG analysis effectively yields two different model solutions (see
Table\,\,\ref{galmodel}). The first one combines quite warm
($T_{kin}=125$ K) and not very dense ($n_{\h2}=10^{3}\,\cc$) molecular
gas with a three times smaller amount of cooler ($T_{kin}=30$ K), much
denser ($n_{\h2}=10^{5}\,\cc$) gas.  The second solution finds
essentially {\it all} molecular gas to be warm ($T_{kin}=150$ K), but
most of it is not very dense ($n_{\h2}=10^{3}\,\cc$) with a small
fraction reaching densities more than an order of magnitude higher
($n_{\h2}=3\times10^{4}\,\cc$).  

However, we have also detected $J$=6-5 $\co$ emission from the center
of NGC~2146, at intensities of about 0.5 to 0.7 times that of the
$J$=4-3 $\co$ emission in the same beam (Figure\,\ref{n2146comp}).
The second of the two model solutions ('all gas is warm') is in much
better agreement with this ratio than the first.  In terms of overall
results, As Table\,\ref{galresult} shows, the difference is not great
in terms of beam-averaged results, as indeed should be expected from
the discussion in Section 4.

\subsection{NGC~3079}

Our $\co$ maps (see Fig.\,\ref{n3079COmaps}) show a just-resolved CO
core superposed on the more extended emission of the edge-on galaxy.
In Fig\,\ref{n3079COmaps}-bottom the rapidly rotating gas clearly
stands out from the more sedately moving gas in the disk of NGC~3079.
The central velocity field is complex.  The dynamics of the central
molecular gas of NGC~3079 have been discussed in considerable detail
by Sofue et al. (2001) and Koda et al. (2002) who propose the presence
of both a weak bar, and an extremely massive black hole in the center.
The diameter of the molecular core in the galaxy plane is consistent
with the diameter of $14''$ derived from millimeter array maps
(e.g. Sofue et al. 2001).  In the $J$=4-3 $\co$ line, the core appears
to be more compact than in the lower transitions, but this just
reflects the contribution from the large-scale disk weakening with
increasing transition.  

%Figure 8: More NGC3079 Maps
\begin{figure}[]
\unitlength1cm
\resizebox{8.5cm}{!}{\rotatebox{270}{\includegraphics*{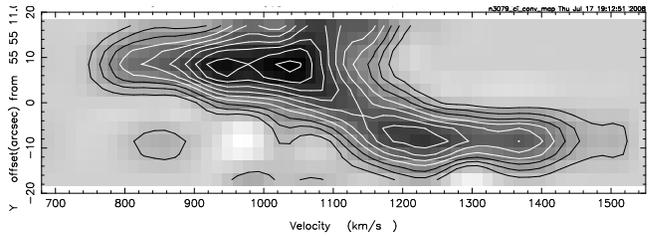}}}
%\begin{minipage}[t]{2.6cm}
%\resizebox{4.5cm}{!}{\rotatebox{270}{\includegraphics*{n3079_ci_8S.ps}}}
%\end{minipage}
%\hfill
%\begin{minipage}[t]{2.6cm}
%\resizebox{4.5cm}{!}{\rotatebox{270}{\includegraphics*{n3079_ci_0.ps}}}
%\end{minipage}
%\hfill
%\begin{minipage}[t]{2.6cm}
%\resizebox{4.5cm}{!}{\rotatebox{270}{\includegraphics*{n3079_ci_8N.ps}}}
%\end{minipage}
\caption[] {Major-axis-velocity map of the [CI] emission from
  NGC~3079. Contour step is 40 mK.  Note relative minimum of
  [CI] emission in the center of NGC~3079. }
\label{n3079ci}
\end{figure}

The behavior of the neutral carbon emission differs from that in
NGC~1068 described above.  In NGC~3079, the [CI] emission does not
peak at the center of the galaxy. Rather, as Fig.\,\ref{n3079ci}
shows, it peaks at either side of the galaxy, and the emission from
the center is relatively weak.  This suggests a distribution in which
a larger fraction of the central region is free of neutral carbon than
of carbon monoxide.  As was the case for NGC~2146, the observed line
ratios allow two essentially different LVG model solutions. The first
one provides a rather good fit to the data and puts a very large
fraction of the molecular gas in a hot ($T_{kin}\,=\,150$ K), tenuous
($n_{\h2}\,=\,10^{2}\,\cc$) gas, leaving a relatively small fraction
of the gas in a cool ($T_{kin}\,=\,30$ K) and very dense
($n_{\h2}\,=\,10^{5}\,\cc$) molecular gas.  The second solution
provides a fit that is not as good, but still within the line ratio
uncertainties.  Now, slightly less than half of the molecular gas is
hot ($T_{kin}\,=\,150$ K) and modestly dense
($n_{\h2}\,=\,10^{3}\,\cc$), whereas the remainder is cold
($T_{kin}\,=\,20$ K) and reasonably dense ($n_{\h2}\,=\,10^{4}\,\cc$).
Braine et al. (1997) found a mean dust temperature of 30 K, which
suggests that the second model solution is more likely than the first.
The mean molecular gas column densities of NGC~3079 are highest of all
the sample galaxies, which is not surprising in view of the almost
exactly edge-on orientation of the galaxy, and the high extinction
($A_{V}\geq 6^{m}$, Israel et al. 1998) towards the very center. The
central molecular mass is $M_{\h2}\,\approx\,2\times10^{8}\,\Msun$,
for either model solution.  This mass is much lower than the values
derived by the various authors quoted above using `standard'
CO-to-$\h2$ conversion factors.  It is, however, very close to the mass
derived by Braine et al. (1997) using a different method.  An
important consequence of this lower molecular gas mass determination
is that the gas no longer constitutes such a large fraction (half!) of
the dynamical mass.

\subsection{NGC~4826}

The maps in Fig.\,\ref{n4826COmaps} show that essentially all
molecular gas is contained with a distance of $30''$ (600 pc) from the
nucleus -- i.e. all molecular gas is located within the radius at which
the rotation curve reverses sign. This is particularly well
illustrated by the $J$=2-1 $\co$ major-axis velocity map at bottom,
Unlike the p-V maps of the other galaxies, this map shows very clear
boundaries at either side.  Although the velocity-integrated CO
emission appears to be peaked on the center in all maps, the
position-velocity maps in Figs.\,\ref{n4826COmaps} and
\ref{n4826moreCO} show that this is not the case. In all transitions
the CO emission reaches a minimum at the center.  This is also seen in
the higher-resolution array maps published by Casoli $\&$ Gerin (1993)
and Garc\'ia-Burillo et al. (2003).

In Fig.\,\ref{n4826COmaps}, the $J$=3-2 $\co$ distribution appears
more compact than that in the $J$=2-1 transition.  This is partly an
effect caused by the significantly higher resolution ($14''$ vs
$21''$) of the former. However, closer analysis shows that the
higher-excited CO gas is indeed more concentrated than the CO in the
ground states.  In all observed transitions, CO emission extends out
to $R\,=\,30''$ (see e.g.  Fig.\,\ref{n4826moreCO}-top), but the
half-width (FWHM) of the CO distribution, corrected for finite
beam width, is $43''$ in the $J$=2-1 transition, and only $20''$ ($R$ =
200 pc) in both the $J$=3-2 and $J$=4-3 transitions
({\it cf} Fig.\,\ref{n4826moreCO}-bottom), as well as in [CI] (not shown).
It thus appears that the hotter molecular gas is preferentially found
closer to the starburst center.

%Figure 9: More NGC4826 Maps
\begin{figure}[]
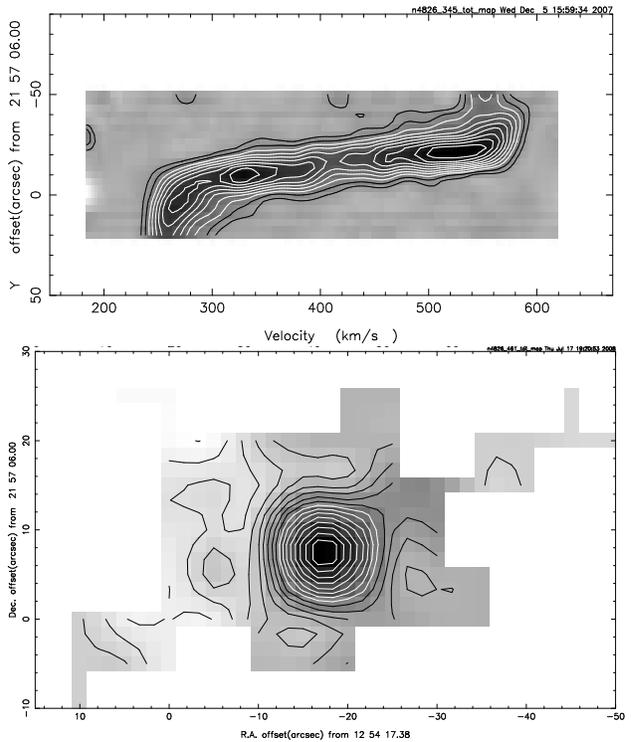

\unitlength1cm
\begin{minipage}[t]{8.5cm}
\resizebox{8.5cm}{!}{\rotatebox{270}{\includegraphics*{N4826_345pV_40mK.ps}}}
\resizebox{8.2cm}{!}{\rotatebox{270}{\includegraphics*{N4826_461_10kkms.ps}}}
\end{minipage}
\caption[] {Top: major-axis-velocity diagram of the $J$=3-2 $\co$
  emission from NGC~4826.  Contour step is 40 mK. Bottom: $J$=4-3
  $\co$ map of NGC~4826 with contour step of 10 $\kkms$.  }
\label{n4826moreCO}
\end{figure}

The $\co$ and $\13co$ line ratios observed in NGC~4826 are not very
well fitted by the model parameters given in Table\,\ref{galmodel}. In
particular the $J$=4-3 $\co$ line is too strong, a situation we also
found in the center of NGC~1068.  The high temperature of this
relatively dense ($n_{\h2}\,\approx\,3000\,\cc$ gas close to the
nucleus opens up the possibility that X-ray emission from the latter
plays a role in the excitation.  A minor radio source (Hummel et
al. 1987; Turner $\&$ Ho 1994) in the center suggests the presence of a
weak AGN (see also Garc\'ia-Burillo et al. 2003; Haan et al. 2008). It
is noteworthy that the molecular gas constitution changes away from
the central peak.  For instance, at a position $18''$ (355 pc) ESE from
the nucleus we find line ratios that suggest either a kinetic
temperature of about 30 K and a density of about 3000 $\cc$, or
comparable amounts of cold, very dense ($T_{kin}\,=10$ K,
$n_{\h2}\,=\,10^{5}\,\cc$) and warm, very tenuous ($T_{kin}\,=60$ K,
$n_{\h2}\,=\,10^{2}\,\cc$) molecular gas.  The lack of more $\13co$
measurements makes it impossible to choose between these alternatives,
but especially the first one is compatible with molecular gas heated
by UV radiation from luminous stars.

\subsection{NGC~7469}

As expected, our maps (Fig.\,\ref{n7469COmaps}) do not show much
detail in this distant galaxy.  Nevertheless, the $J$=3-2 $\co$ maps
show emission from the center superposed on a more extended disk.
Unfortunately, our $J$=4-3 $\co$ map is incomplete in the direction of
the $J$=3-2 extension.  Our angular resolution is insufficient to
resolve the rotation curve of NGC~7469, but once again it can be
clearly seen that the nucleus itself is at a relative CO minimum, as
confirmed by the array maps obtained by Davies et al. (2004).  We may
thus expect the molecular gas parameters to be dominated by the
extended starburst region rather than by the immediate surroundings of
the AGN.  The observed line ratios are well-fitted by the model
parameters listed in Table\,\ref{galmodel}.  These imply that a small
amount (typically a few times $10^{7}\,\Msun$) of the molecular gas is
modestly warm ($T_{kin}$ = 30-60 K) and rather dense
($n_{\h2}\,=\,10^{5}\,\cc$).  It is tempting to associate this
relatively small amount with the AGN-irradiated gas also seen in
near-infrared $\h2$ emission (Genzel et al. 1995; Davies et al. 2004).
The bulk of the molecular gas has mean temperatures ($T_{kin}$ = 1-30
K and mean densities ($n_{\h2}\,=\,10^{2}-10^{3}\,\cc$) appropriate to
extended star formation in the disk and spiral arms.

\subsection{Column densities and masses}

Characteristically, we find $\h2$ column densities averaged over a
$21''$ beam of roughly 1-3 $\times\,10^{21}\,\cm2$, corresponding
to {\it total hydrogen} columns of 2-6 $\times\,10^{21}\,\cm2$.  These
total columns also include the ambient contribution by neutral atomic
hydrogen which we have so far neglected.  HI absorption column
densities actually measured towards these galaxies often exceed the
$\h2$ column densities listed in Table\,\ref{galresult}.  Towards
NGC~1068 HI column densities $N(HI)$ = (1-2) $\times10^{21}\,\cm2$
(Brinks et al. 1997) are seen, but towards NGC~2146 peaks are up to
six times higher than our average $\h2$ value, and reach as high as
$N(HI)$ = (5-12) $\times10^{21}\,\cm2$ (Tarchi et al. 2002) in a
$2.1''\times1.5''$ beam. Towards the nucleus of the edge-on galaxy
NGC~3079, Pedlar et al. (1996) even find $N(HI)\,=
\,27\times10^{21}\,\cm2$ in a $4''\times3''$ beam.  Compared to this,
the mostly face-on galaxy NGC~7469 has rather moderate neutral atomic
hydrogen column densities $N(HI)\,\approx\,3\times10^{21}\,\cm2$),
seen directly in absorption against the nucleus (Beswick et al. 2002)
and also implied by X-ray absorption (Barr, 1986).  Especially in
edge-on galaxies, not all HI seen in absorption necessarily resides in
the central region we have sampled.  Moreover, in all these cases the
effective absorbing area is much smaller (by factors of 30 to 150)
than the large $21''$ beam over which we have averaged our $\h2$ in
Table\,\ref{galresult}.  Our models imply CO, hence $\h2$, surface
filling factors of 5-8 for NGC~2146, NGC~4826, and NGC~7469, of 15 for
NGC~1068, and 33 for NGC~3079.  This suggests that the HI contribution
to the total hydrogen column densities implied by the models is very
small for NGC~1068 (on the order of a few per cent), and somewhat more
(10-30 per cent) for NGC~2146, NGC~3079 and NGC~7469.  In view of the
uncertainty of a factor of two inherent in the model values, the error
introduced by neglecting the HI contribution is negligible.

Our $\h2$ masses are much lower than those calculated previously by
authors who used a `standard' CO-to-$\h2$ conversion factor $X$.  For
instance, in the center of NGC~2146 we find a molecular hydrogen mass
$M_{\h2}\,\approx\,2\times10^{8}\,\Msun$, a factor of 23 lower than
the mass listed by Greve et al. (2006).  Even taking into account that
these authors used the rather high conversion
($X\,=\,3\times10^{20}\,\cm2/(\kkms)$), there is still an order of
magnitude difference with our result.  In the galaxy centers discussed
here, we find $X$ = 0.1-0.2 $\times\,10^{20}\,\cm2/(\kkms)$ which is a
factor of 10-20 lower than the `standard' Milky Way $X$ factors (see
also Strong et al. 2004), but quite similar to those found in other
galaxy centers (see Israel $\&$ Baas 2003 and references therein).
Clearly, $X$ factors derived for molecular clouds in the Solar
Neighborhood do not apply to molecular gas in the very different
environment posed by galaxy centers.  We also note that the present
results are consistent with the low values derived independently for
the same galaxy centers by other authors from e.g. analysis of
infrared observations ({\it cf} Genzel et al. 1995; Braine et al. 1997;
Papadopoulos $\&$ Allen, 2000; Usero et al. 2004; Davies et al. 2004).
Using the 'standard' value of $X$, authors in the past have concluded
to very high gas mass fractions (up to half of the dynamical mass) in
the centers of galaxies.  With the revised low values, there is no
longer a need to assume such extreme conditions. In fact, in our
galaxies we find much more modest mass ratios
$M_{\h2}/M_{dyn}\,\approx\,0.5-1.0\,\%$.

\subsection{Temperature and excitation}

Although a classical UV-illuminated photon-dominated region (PDR)
associated with a star-forming region may reach quite high kinetic
temperatures, the amount of gas at such elevated temperatures is
rather small and limited to the immediate vicinity of the stellar
heating source.  It is very difficult, even in intense star-forming
regions, to reach globally averaged temperatures in excess of 20-30
K.  An example will illustrate this.  If the heating is provided by
luminous ($L\,=\,0.5-2.6\times10^{6}\,\Lsun$. Vacca et al. 1996) O5
stars, and perfect temperature equilibrium is assumed, the separation
between star and heated gas should be about 1.5-2.5 pc at 20 K, and
only 0.6-1.0 pc at 30 K.  

When we look at Table\,\ref{galmodel} we see that only the molecular
line emission from the clear {\it Seyfert} galaxies NGC~1068 and
NGC~7469 may arise from molecular gas dominated by a {\it starburst}.
In the previous section, we have seen that in either galaxy our
resolution was insufficient to separate the emission from the
circumnuclear molecular gas from that of the much more extended
molecular gas associated with the more starburst.  We may therefore
expect that observations similar to those presented in this paper, but
conducted at substantially higher resolutions, might indeed require a
different excitation for the gas close to the nucleus.  As also noted
above, there are already indications that this so in NGC~1068.  In
this respect, millimeter-array observations of this galaxy in high-$J$
transitions of both $\co$ and $\13co$ (SMA, ALMA) would be quite
interesting.

Much if not all of the molecular gas in NGC~2146, which appears to be a
{\it starburst-dominated} galaxy, is in fact far too hot
($T_{kin}$ = 125-150 K) to be excited by UV photons from luminous
stars.  The situation is more ambiguous in the merger galaxy NGC~4826,
where two thirds of the mass is probably heated in PDR fashion, but
one third of the gas is again too hot with $T_{kin}\,\approx$ 60 K.
In NGC~3079, $20-60\%$ of the gas may be heated by the starburst that
is present at a few hundred parsec from the nucleus, but the remaining
relatively diffuse gas is far too hot with a temperature as high as
$T_{kin}\,\approx$ 150 K.  It is noteworthy that the two galaxies
with the hottest gas (NGC~2146 and NGC~3079) are both associated with
a wind-driven X-ray outflow (Armus et al. 1995; Della Ceca et al. 1999;
Pietsch et al. 1998).  One might speculate that the hot molecular gas
component is efficiently heated by the X-rays from these flows (see
Meijerink et al. 2006, 2007) rather than by the UV photons from the
starburst.  

\section{Conclusions}

\begin{enumerate}
\item We have observed the centers of five galaxies with significant
  central activity in the first four transitions of $\co$ and the
  first three transitions of $\13co$, and three of them also in the
  lower transition of [CI].
\item All galaxies were mapped over at least $1'\times1'$ in the
  $J$=2-1 and $J$=3-2 and over smaller areas in the $J$=4-3
  transitions of $\co$.
\item The line ratios, reduced to a standard $21''$ beam vary
  significantly in the sample galaxies. In general, the $J$=4-3 line
  is still relatively strong, indicating elevated temperatures for
  the molecular gas. 
\item Optical depths in $\co$ appear not to be very high, with
  $\co/\13co$ ratios generally in the range of 11-16.  Exceptions are
  NGC~4826 and NGC~7469.  The first, a merger galaxy, has isotopic
  ratios in the range 6-8 suggesting somewhat higher optical depths
  reminiscent of extended molecular clouds in the Solar Neighborhood.
  On the other hand, the distant Seyfert galaxy NGC~7469 has isotopic
  ratios in the range of 16-20, suggesting that much of the gas in
  this galaxy has a low optical depth characteristic of relatively
  tenuous gas.
\item The molecular gas is present in the form of clearly identifiable
  central condensations of outer radius $R=0.3-1.5$ kpc. In all
  galaxies, the molecular gas distribution exhibits a local minimum at
  the nucleus itself.
\item In all galaxy central regions, we find moderate $\h2$ masses of
  $1-3\times10^{8}\,\Msun$, except in the nearby NGC~4826 where the
  mass is no more than $3\times10^{7}\,\Msun$. These masses are
  typically about one per cent of the dynamical mass in the
  same region.  The CO-to-$\h2$ conversion factors $X$ implied by
  these masses and the observed $J$=1-0 $\co$ line intensities is
  typically an order of magnitude lower than the `standard' Solar
  Neighborhood $X$-factor.
\item Intriguingly, the excitation of molecular gas in the Seyfert
  galaxies NGC~1068 and NGC~7469 appears to be dominated by UV-photons
  from the extended circumnuclear starburst. In contrast, the
  molecular gas in the starburst galaxy NGC~2146, and part of the
  molecular gas in NGC~3079 (starburst cum AGN) are probably excited
  by Xrays from their massive wind-driven outflows. The situation for
  NGC~4826 is less clear.
\end{enumerate}

\acknowledgements{I am much indebted to the late Fred Baas, who died
  suddenly and far too early in 2001, for carefully planning and
  performing many of the observations described in this paper.  I also
  thank the various JCMT observers who made additional observations
  in queue mode.  Finally, it is always a pleasure to thank the
  personnel of both the JCMT and the IRAM 30m telescope for their able
  and generous support.  }

\end{document}